\documentclass[11pt,a4paper]{article}

\usepackage[
    a4paper,
    left=2.2cm,
    right=2.2cm,
    top=2.5cm,
    bottom=2.5cm
]{geometry}

\usepackage{graphicx}
\newcommand{\figwidth}[1]{\dimexpr\linewidth*#1/100\relax}
\usepackage{amsmath}
\usepackage{amssymb}
\usepackage{float}
\usepackage{multirow}
\usepackage{authblk}
\usepackage{hyperref}

\hypersetup{
    hidelinks
}

\title{Quantum Anomaly Detection in network traffic using Variational Quantum Classifiers}

\author[4]{Mir H M Ilham}
\author[2,5]{Hardique Dasore}
\author[3]{Abdulrahman Ababneh}
\author[3]{Hafidz A Kabira}
\author[1]{Samyam Shrestha}
\author[3]{Jeyasri Sundar}
\author[2]{Shah Haque}
\author[1,2]{Eric Howard}

\affil[1]{School of Mathematical and Physical Sciences, Macquarie University, Sydney, NSW, Australia}
\affil[2]{Southern Cross Institute, School of Computer Science, Sydney, Australia}
\affil[3]{University of Sydney, Sydney, NSW, Australia}
\affil[4]{East Delta University, Chattogram, Bangladesh}
\affil[5]{School of Computer Science and Engineering, SR University, Warangal, Telangana, India}

\begin{document}

\maketitle
\begin{abstract}
The increasingly complex cybersecurity landscape requires more advanced network anomaly detection systems. Although, classical machine learning models are known to be effective, they reach computational bottlenecks when processing high-dimensional data. This study investigates the potential of Quantum Machine Learning (QML) for network traffic anomaly detection by evaluating four distinct Variational Quantum Classifier (VQC) architectures. A synthetic network traffic dataset is generated for the evaluations. The VQC models systematically varied feature encoding methods, parameterized ansatz circuits, quantum neural network types, and classical optimizers. Experimental results show that a trade-off exists between circuit expressivity and optimization stability. Deeper circuits (VQC-4) achieved higher accuracy of upto 89.0\%, while simultaneously being sensitive to training conditions. A shallower variant (VQC-1) under performed, although it had consistent and stable optimization curves. Class-dependent biases were also observed such as in VQC-2, where the model struggled to find an effective decision boundary, and produced a large number of false negatives. The best performing model in this study (VQC-3) combined ZFeatureMap with RealAmplitudes ansatz, optimized using ADAM. Comparative analysis against classical machine learning models such as Support Vector Machine and Random Forest reveal a performance gap, where classical models are able to achieve perfect classification performance on the synthetic data. It is concluded that classical algorithms remain dominant in their ability to classify structured data with relatively simple class separations. However, due to the observed potential of certain VQC configurations to learn decision boundaries to classify data, their utility for future applications in noisier, higher-dimensional problems is highlighted.

\vspace{0.2 cm}

\textit{keywords:} Variational Quantum Classifier, Quantum Machine Learning, Network Anomaly Detection, NISQ Computing, Hybrid Quantum–Classical Models, Cybersecurity
\end{abstract}

\section{Introduction}

As technology evolves, the digital landscape faces an unprecedented increase in cybersecurity threats. Approximately 4,000 cyber-attacks are carried out daily on network systems \cite{fortinet2025cybersecuritystats}. The average cost of a data breach globally is over \$4.88 million. Attacks and threats continue to evolve with the introduction of zero-day exploits and other persistent and evasive threats \cite{ibm2024databreach}. Traditional security controls to protect an organization may not be enough to mitigate the residual risk, given the complex nature of evolving threats. This creates an imperative need to deploy novel approaches to security and forensics to detect evasive threats.

A critical component of modern day cybersecurity is the use of network anomaly detection systems \cite{garcia2009anomaly}. Unlike traditional signature based detection approaches which are effective in detecting known threats, network anomaly detection systems can leverage ML and AI to extend detection to previously unknown threats \cite{apruzzese2018effectiveness}. Emerging technologies such as deep learning and ML 2.0 are powerful to address the challenges faced in modeling complex and non-linear relationships in network traffic data \cite{buczak2015survey}. The use of ML, however, needs to consider challenges such as feature extraction from large data sets, class imbalance, high dimensional feature spaces, high computation demands, and real time traffic analysis \cite{ferrag2020deep}. In addition, ML systems are vulnerable to adversarial attacks, which further motivates the need for robust detection methodologies \cite{apruzzese2019addressing, papernot2017practical}.

Quantum computing offers a distinct computational paradigm that takes advantage of quantum phenomena such as superposition, entanglement, and interference. It has demonstrated computational advantages for specific tasks and there is ongoing investigations regarding its broader algorithmic applications \cite{preskill2018quantum, arute2019quantum}. At present the available hardware is limited, non-fault tolerant, has imperfect gate fidelities, restricted connectivity, and finite coherence times \cite{bharti2022noisy, xie2025advances}. The term Noisy Intermediate-Scale Quantum (NISQ) is used when describing current quantum devices. Near term applications of quantum computing often exploit hybrid quantum-classical algorithms where parameterized quantum circuits are optimized using classical optimization \cite{cerezo2021variational, peruzzo2014variational}. As a consequence, Quantum Machine Learning (QML) has emerged as an active research area, where representation in quantum feature space, and variational quantum algorithms offer promising solutions for future applications \cite{dunjko2018machine, schuld2019quantum}.

Variational Quantum Classifiers (VQC) represent a QML approach that is suitable for the NISQ era. They are composed of parameterized quantum circuits optimized by classical algorithms with the objective of minimizing an error function \cite{havlivcek2019supervised, schuld2020circuit}. In this study, a structured evaluation of multiple VQC architectures is carried out by varying various components of the training pipeline including feature encoding methods, ansatz structures, circuit depths, Quantum Neural Network (QNN) types, and classical optimizers.

The primary contributions of this work are as follows.
\begin{itemize}
    \item Four distinct VQC architectures are comprehensively evaluated. Each architecture is varied by using combinations of different feature maps, ansatz structures, QNN types, and classical optimizers.
    \item To observe the performance of VQCs under variations in training parameters such as random initialization, dataset size, class imbalance ratios, train-test splits, and optimizer iterations, a sensitivity analysis study is conducted.This work has insights into the stability of the models developed and their ability to generalize.
    \item It benchmarks quantum classifiers against some classical machine learning baselines. With these baselines, it sets realistic expectations for performance of quantum algorithms of the NISQ era.
    \item Discussions are provided regarding the trade-offs between circuit depth, the complexity of the optimization functions, and the performance of the models.
    \item Model performance and misclassifications are analyzed by reviewing error distributions.
\end{itemize}

The rest of this paper is organized as follows: Section 2 reviews related works in VQCs and quantum machine learning in cybersecurity. Section 3 describes the methodology. Section 4 presents the experimental results and analysis, and Section 5 summarizes the key findings, and gives future work directions.

\section{Related Works}

Quantum Machine Learning (QML) applications in the domain of cybersecurity and anomaly detection is an increasingly growing field built on the basis of variational quantum algorithms. Maheshwari et al. \cite{maheshwari2021vqc} investigated the effects of preprocessing and quantum data encoding on VQC performance. Their amplitude encoding implementation achieved 98.40\%, 67.3\%, and 74.5\% accuracy in synthetic, sonar, and diabetes datasets, respectively. Compared to their baseline VQC, the proposed approach improved performance on the synthetic and diabetes datasets while giving a lower accuracy for the sonar dataset. The study highlights that VQC performance is strongly dependent on preprocessing and quantum data encoding.

VQCs have also been examined for medical applications. Khan et al. \cite{khan2025diabetes} varied qubit counts, feature maps, and optimizers among COBYLA, SPSA, SLSQP and studied the performance across 76 variants for diabetes prediction. Their best VQC setup used ZZ feature map with COBYLA optimizer and achieved an accuracy of 85\%, outperforming several classical baselines such as SVM, decision tree, and random forest. Likewise, for chronic kidney disease identification a framework integrating kalman filtering was proposed by Parthasarathi et al. \cite{parthasarathi2025ckd}. It leverages sparse autoencoders for handling class imbalance, and LASSO feature selection in the preprocessing stage. They achieved an accuracy of 99.2\% on the UCI CKD dataset, thus underscoring the potential of hybrid quantum-classical models for medical decision support.

Kavitha et al. \cite{ss2025internet} evaluated Quantum Support Vector Classifier and Quantum Random Forest models against their classical counterparts for network intrusion detection. Experiments were conducted using the UNSW-NB15 dataset, where optimized quantum configurations attained improved classification performance. They identified simulator dependence as a limitation, and interpretability of results as an important consideration for deployment in real cybersecurity applications. The ``QuantumNetSec'' system introduced by Abreu et al. \cite{abreu2025quantumnetsec} used backend aware circuit optimization to mitigate noise. It achieved strong QML performance across network-security benchmarks including UNSW-NB15 and CIC-IDS17. Kalinin and Krundyshev \cite{kalinin2023intrusion} reported an accuracy of 98\% and faster processing in their experimental setting for QML-based intrusion detection. However, they also highlight that current hardware limitations prevent large-scale deployment. Complementing classifier-based approaches, Mridha et al. \cite{mridha2025arxiv} proposed a quantum Bayesian inference framework for intrusion detection in which prior and conditional probabilities are encoded in quantum circuits and posterior distributions are obtained through statevector simulation and post-selection, providing an interpretable quantum-native approach to probabilistic reasoning in cybersecurity.

For anomaly detection, Kumar et al. \cite{kumar2025anomaly} utilized the idea of quantum-amplified feature representation and fidelity-based distance metrics. Their method used a hybrid of quantum support vector machines (QSVMs) and quantum k-means clustering. Performance improved for imbalanced and non-linear cases by mapping features into high dimensional states. For NSL-KDD, Rahman et al. \cite{rahman2024vqc} reported an accuracy of 90 percent by fine-tuning VQCs using the EfficientSU2 ansatz and COBYLA optimizer. The results indicated that the performance of the circuit architecture influenced the results, and using PCA to reduce features improved the results.

Advancements in Quantum Deep Learning (QDL) have enabled the use of hybrid QDL architectures for various cybersecurity relevant scenarios. To enhance IoT network security Hdaib et al. \cite{hdaib2024quantumdl} have integrated quantum autoencoders for feature compression with algorithms such as quantum k-nearest neighbor. Rosa-Remedios and Caballero-Gil \cite{rosa2025optimizing} have also proposed systematic methodologies for cybersecurity tasks like phishing detection, malware classification, and spam detection. They looked into different types of feature mapping and variation circuit anstazes, and contributed structured framework for parameter tuning and evaluation. The task of malicious URL detection have been studied by Eze et al. \cite{eze2025url}. They demonstrate that while QSVM outperforms classical SVM consistently, classical CNNs still surpass QCNNs due to current hardware limitations.

Existing literature highlight the potential of quantum machine learning for cybersecurity applications. However, crucial gaps exist in understanding quantum classifiers for anomaly detection tasks. Majority of the studies focus on evaluating individual circuit configurations or specific strategies for optimization. A systematic comparative analysis of how architectural choices have an effect on classification performance and training stability is largely absent. In addition, the robustness of VQC models under varying training conditions has not been investigated thoroughly. This research study addresses the gap in the current literature by structured evaluation and sensitivity analysis of multiple VQC architectures obtained by varying feature encoding methods, ansatz structures, circuit depths, quantum neural network types, and classical optimizers.

\section{Methodology}

\subsection{Problem Formulation}

Our research methodology formulates anomaly detection on synthetic network traffic data as a supervised binary classification problem. The data points are features vectors of real-numbers that represent fixed time snapshots of network behavior. Features include traffic intensity, reliability, and protocol type. The objective is for the classifier to distinguish between normal and anomalous traffic based on deviations from usual statistical pattern.

Formally, let \( x \in \mathbb{R}^4 \) denote a feature vector of four numerical attributes. Let the class label be denoted by \( y \in \{0,1\} \), where \( y = 0 \) is the label for normal and \( y = 1 \) is the label for anomalous. The objective is to learn a decision function given a training set \( \{(x_i, y_i)\}_{i=1}^N \).

\[
f : \mathbb{R}^4 \rightarrow [0,1]
\]

The function should output a confidence score with a binary prediction.

\subsection{Overview of Methodology}

We use a hybrid quantum-classical training pipeline for this study. We focus on Variational Quantum Classifier (VQC) architectures to study the binary classification capability of parameterized quantum circuits.

\begin{figure}[H]
\centering
\includegraphics[width=\figwidth{75}]{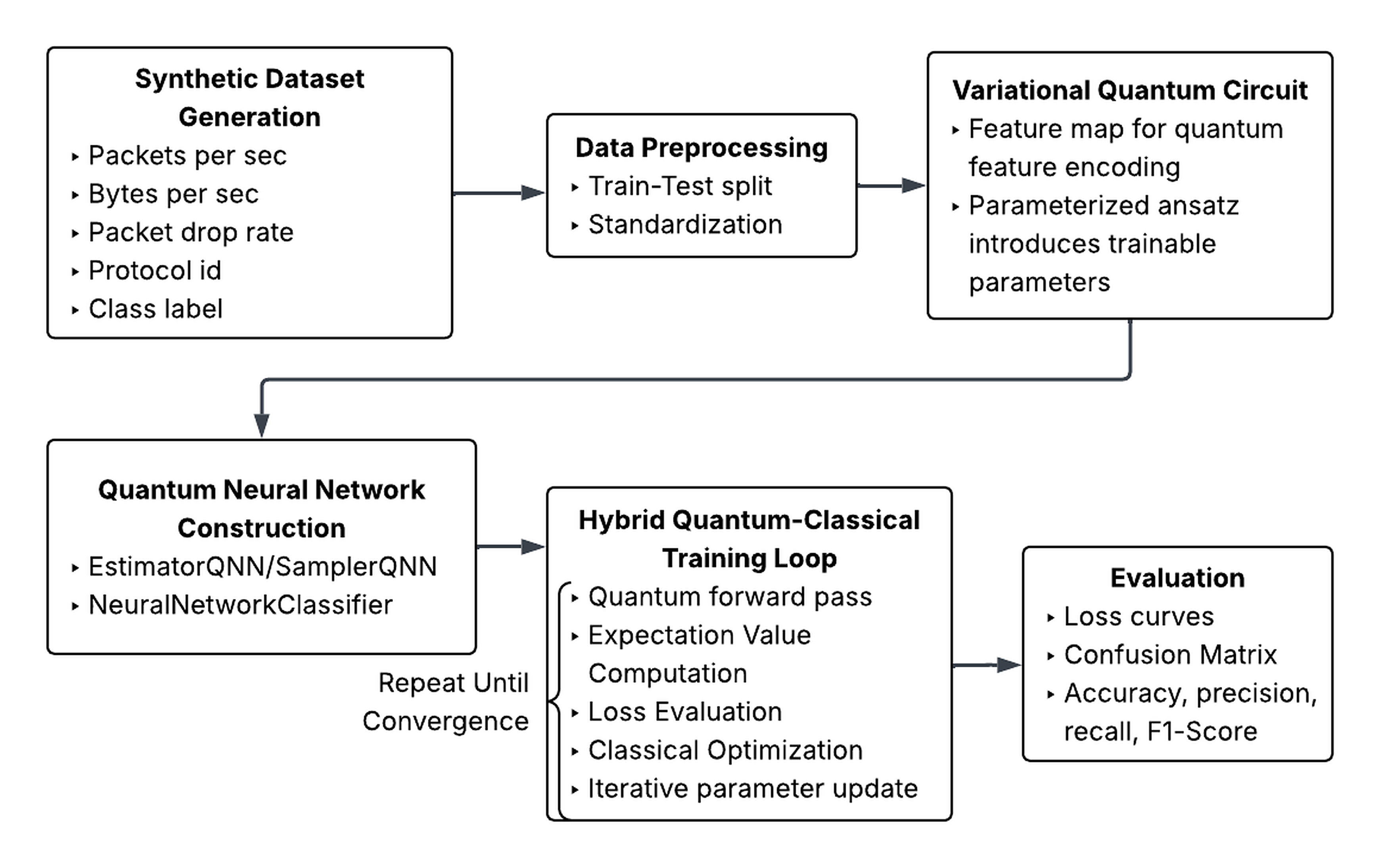}
\caption{The methodology overview illustrates the pipeline steps of data generation, preprocessing, quantum feature encoding, hybrid quantum-classical training, and model evaluation.}
\label{fig:methodology-overview}
\end{figure}

The workflow shown in Figure~\ref{fig:methodology-overview} begins with generation of the synthetic network traffic dataset. It is then split into train and test subsets. To ensure consistent feature scales and stable optimization during model training, classical feature standardization is applied.

Quantum Feature Maps take encoded feature vectors and transform them into quantum states. Parameterized ansatz circuits act on these quantum states. The interplay of these two ingredients, known as Variational Quantum Circuits (VQC), enables learning of data patterns. Through the use of different types of feature maps and ansatz structures, we designed four different VQC architectures.

The variational circuits are incorporated into the architecture of a Quantum Neural Network (QNN) and trained via a quantum-classical hybrid loop. During the training, quantum forward passes are made to produce expectation values. These expectation values are then used to calculate the classification loss. Classical optimizers, acting on variational circuit parameters, work to minimize the loss value. This process is repeated until the loss converges.

For the final assessment of our model, we used accuracy, precision, and F1 score. Additionally, we used loss curves and plotted confusion matrices. The analysis of these measures gave us insights about the models behavior and classification performance.

\subsection{Dataset Generation}

\subsubsection{Synthetic Network Traffic Data Modeling}

A synthetic network traffic dataset was generated and populated with numerical features that describe various network traffic metrics (i.e. packets/second, bytes/second, packet drop rate, and protocol type).

Data points were drawn from statistical distributions in order to create both normal and anomalous data behaviors. Different distributions were used for each, in order to replicate naturally occurring differences. For continuous features, anomalous samples were drawn using a Gaussian distribution with mean and variance different from those used for normal samples. This created statistical patterns that, when learned, allow for the classification of normal and anomalous samples by the Variational Quantum Classifier.

Data for packets per second in normal and anomalous traffic were modeled in the following ways:

Normal traffic:
\[
P_{normal} \sim \mathcal{N}(100, 10^2)
\]
Anomalous traffic:
\[
P_{anomaly} \sim \mathcal{N}(300, 40^2)
\]

Similarly, packet drop rates were modeled using separate Gaussian distributions:

Normal traffic:
\[
D_{normal} \sim \mathcal{N}(0.01, 0.005^2)
\]
Anomalous traffic:
\[
D_{anomaly} \sim \mathcal{N}(0.1, 0.03^2)
\]

The bytes-per-second feature was generated as a multiplicative function of packet transmission rate and an additional Gaussian scaling factor:

Normal traffic:
\[
B = P \times \mathcal{N}(10, 2^2)
\]
Anomalous traffic:
\[
B = P \times \mathcal{N}(15, 5^2)
\]

The protocol identifier is a label encoded value where $1$ denotes TCP, $2$ denotes UDP, $3$ denotes ICMP.

\begin{table}[h]
\centering
\renewcommand{\arraystretch}{1.2}
% \resizebox{\linewidth}{!}{
\makebox[\textwidth][c]{
\begin{tabular}{lcc}
\hline
Feature & Normal Traffic Distribution & Anomalous Traffic Distribution \\
\hline
Packets per second & $\mathcal{N}(100, 10^2)$ & $\mathcal{N}(300, 40^2)$ \\
Bytes per second & $P \times \mathcal{N}(10, 2^2)$ & $P \times \mathcal{N}(15, 5^2)$ \\
Packet drop rate & $\mathcal{N}(0.01, 0.005^2)$ & $\mathcal{N}(0.1, 0.03^2)$ \\
Protocol identifier & TCP (1) & UDP/ICMP (2, 3) \\
\hline 
\end{tabular} 
}
\caption{Summary of generated feature distributions for both categories}
\label{tab:summary-of-gen-features}
\end{table}

\subsubsection{Class Distribution and Labeling}
The dataset was generated by keeping the anomaly ratio fixed at 50\% of the total samples. Let N denote the total number of generated samples and r represent the anomaly ratio. The number of anomalous and normal samples can therefore be expressed as:
\[
N_{anomaly} = rN
\]
\[
N_{normal} = (1-r)N
\]

Normal traffic samples were assigned the label 0, while anomalous traffic samples were assigned the label 1.

\subsection{Exploratory Data Analysis}

\begin{figure}[H]
\centering
\includegraphics[width=\figwidth{80}]{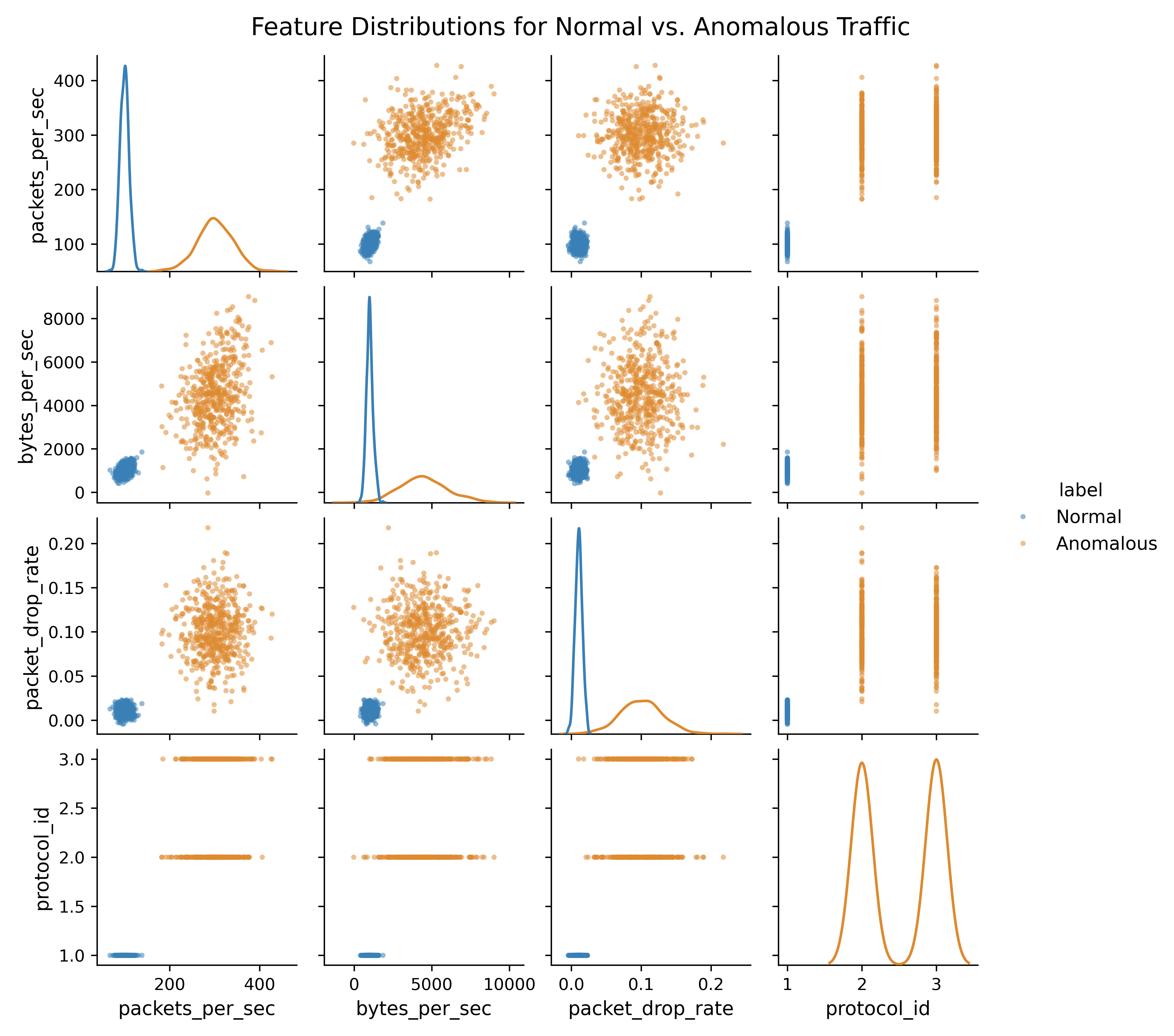}
\caption{The feature distribution plot visualizes both individual feature distributions (diagonal subplots) and pairwise relationships between features in a grid format.}
\label{fig:dataset-pairplot}
\end{figure}

An analysis of the dataset was conducted by visualizing feature distributions, and PCA based  dimensionality reduction. The plots provide insights regarding class separation, and correlation among attributes.

The feature distribution plot in Figure~\ref{fig:dataset-pairplot} shows the distribution of a feature along the diagonal subplots. Off-diagonal subplots show the relationship between variables in a scatter plot. The different classes are color encoded. Normal instances form compact clusters around lower packet rates, byte volumes, and packet drop rates. This is expected because normal traffic would ideally have low variance and stable behavior. In contrast, anomalous data points are widely dispersed, with significantly higher packet rates, byte volumes, and packet drop rates. The spread of the anomalous data cluster is also greater. The protocol feature cleanly separates both classes, where normal traffic is concentrated at protocol 1, and anomalies are distributed across protocols 2 and 3.

\begin{figure}[H]
\centering
\includegraphics[width=\figwidth{75}]{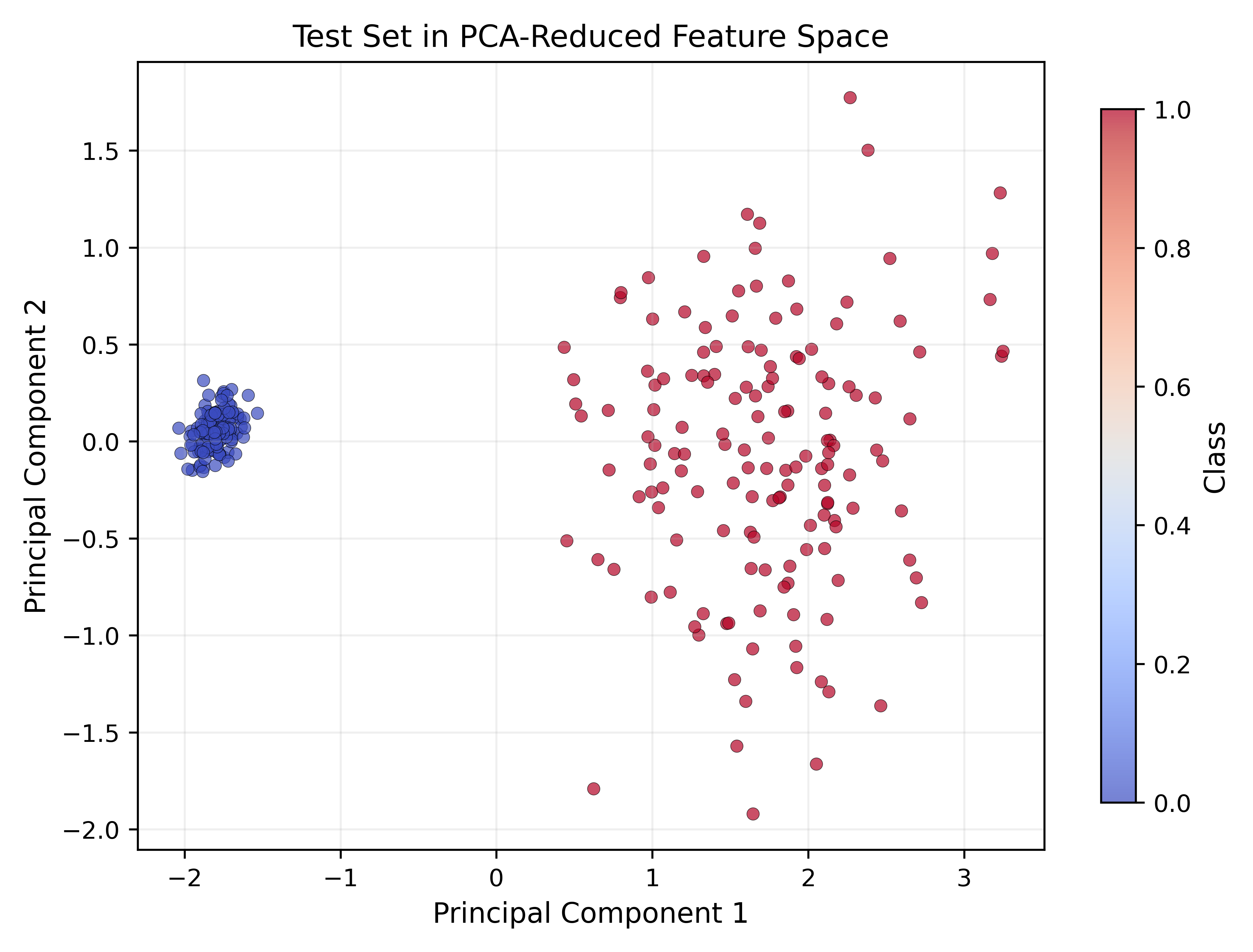}
\caption{Principal Component Analysis (PCA) was applied to project the dataset features into a two dimensional space. It is a statistical dimensionality reduction technique that transforms the data while preserving maximum possible variance.}
\label{fig:PCA}
\end{figure}

From the PCA visualization in Figure~\ref{fig:PCA}, it can be observed that there is a clear and well defined separation between the normal and anomalous class. Normal samples form a compact cluster, and anomalous samples are widely dispersed but remain distinctly separated.

\begin{figure}[H]
\centering
\includegraphics[width=\figwidth{65}]{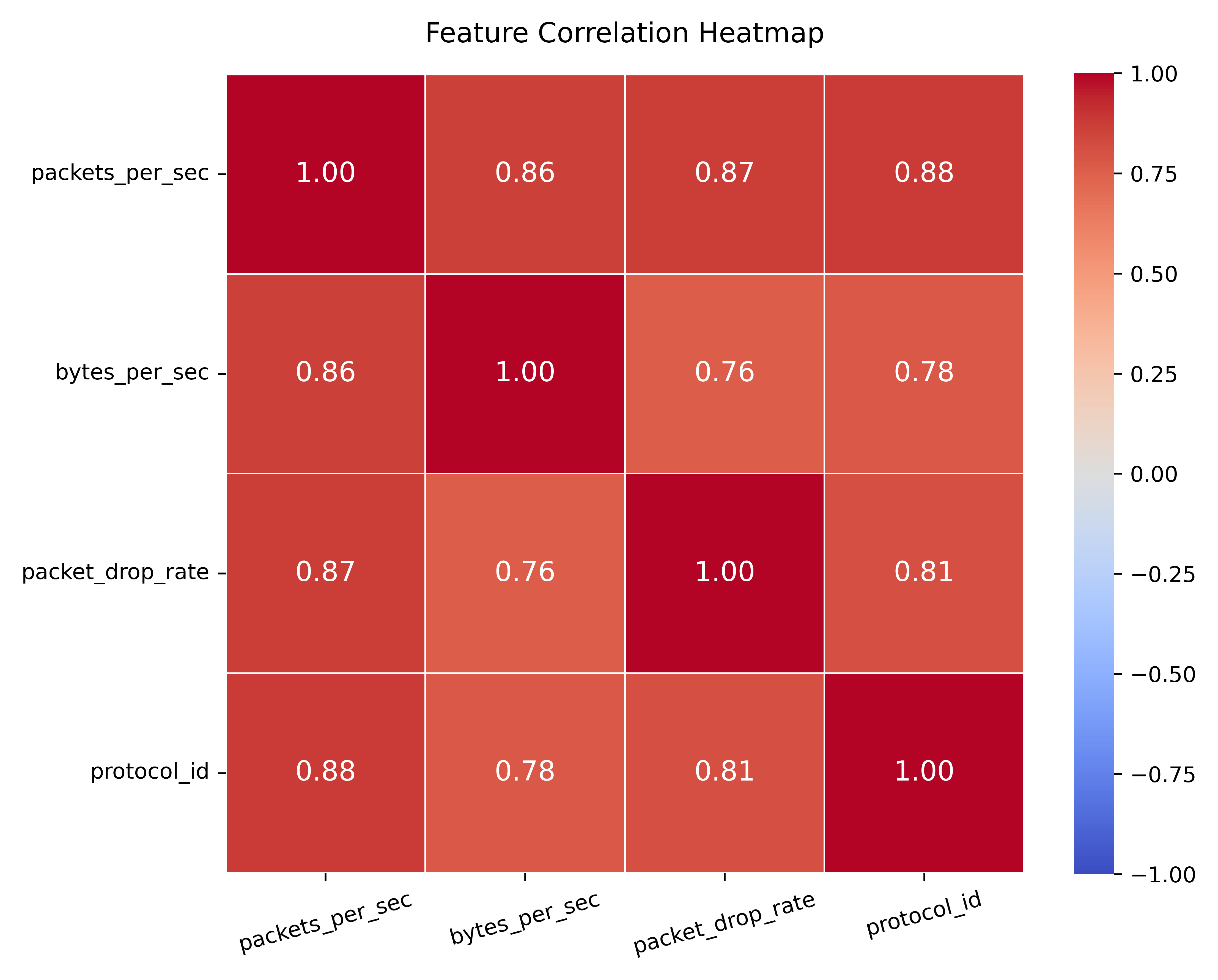}
\caption{A feature correlation heatmap visualizes correlation among variables in the data. In this case, it can be observed that strong positive relationships exist among all features.}
\label{fig:correlation-heatmap}
\end{figure}

The heatmap in Figure~\ref{fig:correlation-heatmap} reveals that $packets\_per\_sec$ is highly correlated with $bytes\_per\_sec$ (0.86), $packet\_drop\_rate$ (0.87), and $protocol\_id$ (0.88). Similar high correlations are observed for the other cases as well. This behavior of network traffic is that an increase of packet transmission will certainly lead to increase of bandwidth consumption and will eventually result in congestion effects such as packet loss.

In terms of networking, this sort of data behavior typically indicates that normal traffic is a result of normal stable TCP communication, where loss is low and throughput is steady. On the other hand, anomalous traffic indicates flooding. High packet drop rate is an indicator of symptoms of congestion, which are often seen in misconfigurations and attacks. Use of UDP/ICMP in this context, serves to strengthen this case, since these protocols are generally used in unbounded connectionless traffic like DDoS attacks and scanning.

\subsection{Data Preprocessing Pipeline}

\subsubsection{Train-Test Splitting}

Before model training, we created a stratified train-test split of the generated dataset to partition it into training and testing sets. We set the evaluation size to 30\% of the data. Stratification was used to keep the balance of normal and anomalous samples during training and evaluation.

\subsubsection{Feature Scaling}

Feature scaling is essential for quantum machine learning because quantum feature encoding circuits are very sensitive to the absolute values of the features. So, after dataset partitioning, feature standardization was performed. It transforms each feature to a zero-mean unit-variance representation, thus eliminating differences in scales among the input features.

The formula for standardization is as follows:

\[
x' = \frac{x - \mu}{\sigma}
\]

where $x$ denotes the original feature value, $\mu$ represents the feature mean, and $\sigma$ denotes the standard deviation.

\subsection{Quantum Model Design}

\subsubsection{Variational Quantum Classifier (VQC)}

A Variational Quantum Classifier is a hybrid quantum-classical algorithm. It translates classical data into a quantum state, manipulates that state using tunable parameters in a quantum circuit, and measures the result to make a prediction. The parameterized quantum circuit is trained using a classical optimizer. This approach enables leveraging quantum state space representations while using classical optimization.

Given an input feature vector $x_i \in \mathbb{R}^d$, where $d$ denotes the feature count, the expectation value of the quantum circuit is calculated according to the following formula:

\[
f(x_i, \theta) = \langle \psi(x_i) | U^\dagger(\theta) \hat{O} U(\theta) | \psi(x_i) \rangle
\]

Reading the right side of the equation from inside out (following standard quantum mechanics convention) the term $|\psi(x_i)\rangle$ represents the classical input vector $x_i \in \mathbb{R}^d$ encoded into a quantum state vector. Common methods include angle encoding or amplitude encoding.

The term $U(\theta)$ is a unitary transformation consisting of a sequence of quantum gates like $RX$, $RY$, and $CNOT$ with adjustable parameters $\theta$ that are usually rotation angles. It serves the same purpose as the weights in a classical neural network layer. When $U(\theta)$ is applied to the encoded data, it evolves the quantum state:

\[
|\phi(x_i, \theta)\rangle = U(\theta)|\psi(x_i)\rangle
\]

The terms $\hat{O}$ is a Hermitian operator representing the physical observable being measured. To extract a classical number from a quantum system, it must be measured.

The remainder terms in the left side $\langle \psi(x_i) | U^\dagger(\theta)$ is the conjugate transpose of the evolved state. Placing the operator $\hat{O}$ between the bra and the ket yields the expected value, which is always a real number.

The final scalar output $f(x_i, \theta)$ serves as the prediction function for the machine learning model. When performing a binary classification task, it can be passed through a threshold. For example, if the output is $\geq 0$, predict Class 1 else if it is $< 0$, predict Class 0. This output is sent back to a classical computer to evaluate a loss function against the true data labels. A classical optimizer such as Adam or SPSA computes the gradients and updates the parameter angles $\theta$ to minimize the loss, repeating the process until the classifier learns the boundary.

For the hybrid optimization loop, the training objective is defined by the following equation:

\[
\mathcal{L}(\theta) = \frac{1}{N} \sum_{i=1}^{N} \ell(y_i, f(x_i, \theta))
\]

Where $\ell(\cdot)$ is the classification loss function,  and $\theta$ represents the trainable parameters.

Different loss functions are used with NeuralNetworkClassifier based on the type of network provided. EstimatorQNN outputs raw, continuous expectation values typically between $[-1, 1]$. This output is treated as a regression-style continuous score, and the classifier defaults to Mean Squared Error. SamplerQNN outputs a probability distribution over classes. As the output is a probability distribution, Cross-Entropy Loss is used.

The optimizer iteratively updates the parameters given by the formula below, where $\eta$ denotes the learning rate:

\[
\theta \leftarrow \theta - \eta \nabla_\theta \mathcal{L}(\theta)
\]

This study employs four VQC models, where each consists of the following: 1) Feature Encoding Circuit (Feature Map), 2) Variational Ansatz (Trainable Circuit), 3) Measurement Layer, 4) Classical Optimization Loop. The following table summarizes the types of components used in each model.

\begin{table}[h]
\centering
% \resizebox{\linewidth}{!}{
\makebox[\textwidth][c]{
\begin{tabular}{lcccc}
\hline
Model & Feature Map & Ansatz & QNN Type & Optimizer \\
\hline
VQC-1 & ZZFeatureMap (r=2) & RealAmplitudes (r=2) & Estimator & SPSA \\
VQC-2 & PauliFeatureMap (r=2) & TwoLocal (r=2) & Sampler & SPSA \\
VQC-3 & ZFeatureMap (r=2) & RealAmplitudes (r=3) & Estimator & ADAM \\
VQC-4 & ZZFeatureMap (r=4) & RealAmplitudes (r=4) & Estimator & COBYLA \\
\hline 
\end{tabular}
}
\caption{Summary of Model Components}
\label{tab:model-component-summary}
\end{table}

The generated dataset used in this study consists of four features. Consequently, all Variational Quantum Classifier architectures were constructed using four qubits in order to maintain direct feature-to-qubit encoding consistency.

\subsubsection{VQC-1: ZZFeatureMap + RealAmplitudes}

VQC-1 represents a shallow variational quantum classifier designed to balance expressivity and trainability. The model employs a structured feature encoding mechanism followed by a low-depth parameterized circuit.

\begin{figure}[H]
\centering
\makebox[\textwidth][c]{
  \includegraphics[width=\figwidth{75}]{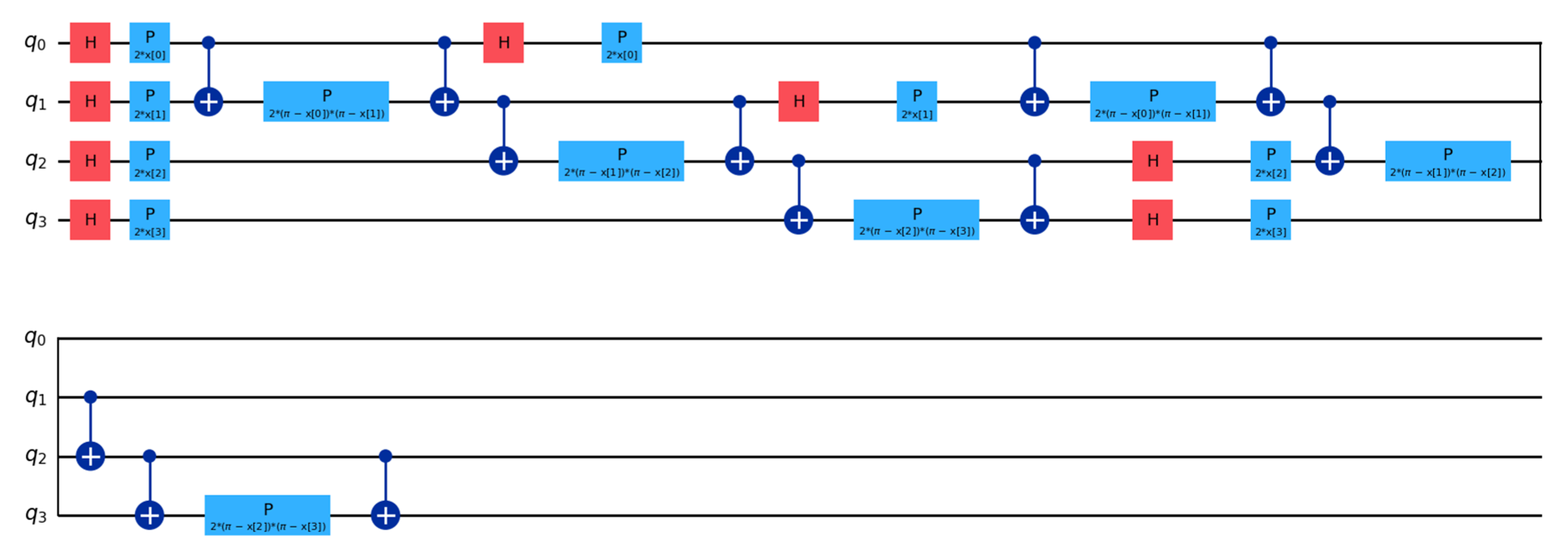}
}
\caption{VQC-1 feature encoding using the ZZFeatureMap maps classical input features into a quantum state by applying both single-qubit rotations and entangling operations.}
\label{fig:vqc1-feature-map}
\end{figure}

\begin{figure}[H]
    \centering
    \includegraphics[width=\figwidth{65}]{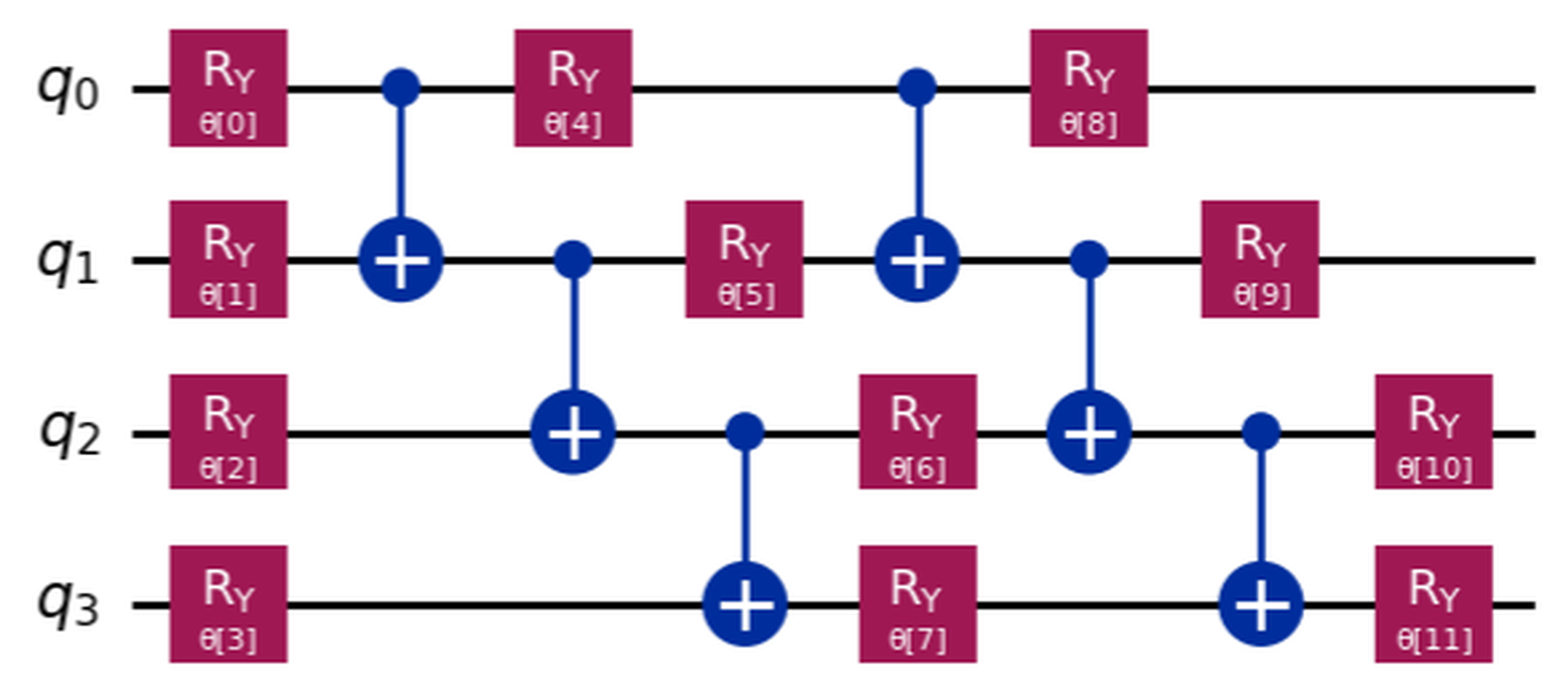}
    \caption{VQC-1 Ansatz consists of parameterized $R_y$ rotation gates interleaved with nearest-neighbor entangling CNOT operations arranged in a shallow repeated structure.}
    \label{fig:vqc1-ansatz}

\end{figure}

The feature encoding process using ZZFeatureMap, illustrated in Figure~\ref{fig:vqc1-feature-map}, can be expressed as:

\[
x \rightarrow |\psi(x)\rangle
\]

The unitary transformation implemented by the ZZFeatureMap is given by the following formula for $d$ repetitions on $n$ qubits:

\[
U_{ZZ}(x) = \left( \exp\left(i \sum_{i<j} \phi_{ij}(x) Z_i Z_j\right) \exp\left(i \sum_{i=1}^n \phi_i(x) Z_i\right) H^{\otimes n} \right)^d
\]

Here, the $Z_i Z_j$ operators introduce pairwise entanglement between qubits, while the phase term $\phi_{ij}(x)$, typically dependent on feature products $x_i x_j$, enables the encoding of non-linear correlations within the data. This makes the feature map particularly suitable for anomaly detection tasks where feature interactions are significant.

Following the encoding stage, a RealAmplitudes ansatz with two repetitions (Figure~\ref{fig:vqc1-ansatz}) is applied. The unitary transformation of the ansatz can be expressed as follows for a circuit with $L$ repetitions:

\[
U_{RA}(\theta)=\left(\prod_{l=1}^{L}\left(\bigotimes_{i=1}^{n}R_{y}(\theta_{i,l})U_{CNOT}\right)\right)\bigotimes_{i=1}^{n}R_{y}(\theta_{i,0})
\]

The circuit is able to learn meaningful representations within a limited depth. Its structure ensures parameters are efficiently used, while maintaining sufficient qubit entanglement.

EstimatorQNN is used as the quantum neural network type. The model output is computed as expectation values of observables. This choice particular choice of QNN is due to the relatively shallow circuit structure, because expectation estimation is computationally efficient and outputs are smoother for optimization.

Model parameters are optimized using the SPSA optimizer. Stochastic perturbations approximates gradients in this method. The optimizer is robust to noise and has low computational overhead, which makes it particularly suitable for quantum settings.

\subsubsection{VQC-2: PauliFeatureMap + TwoLocal Ansatz}

VQC-2 uses a more generalized feature encoding method and a flexible ansatz structure. The aim is to extend representational capacity over the latter model.

\begin{figure}[H]
    \centering
    \makebox[\textwidth][c]{
        \includegraphics[width=\figwidth{75}]{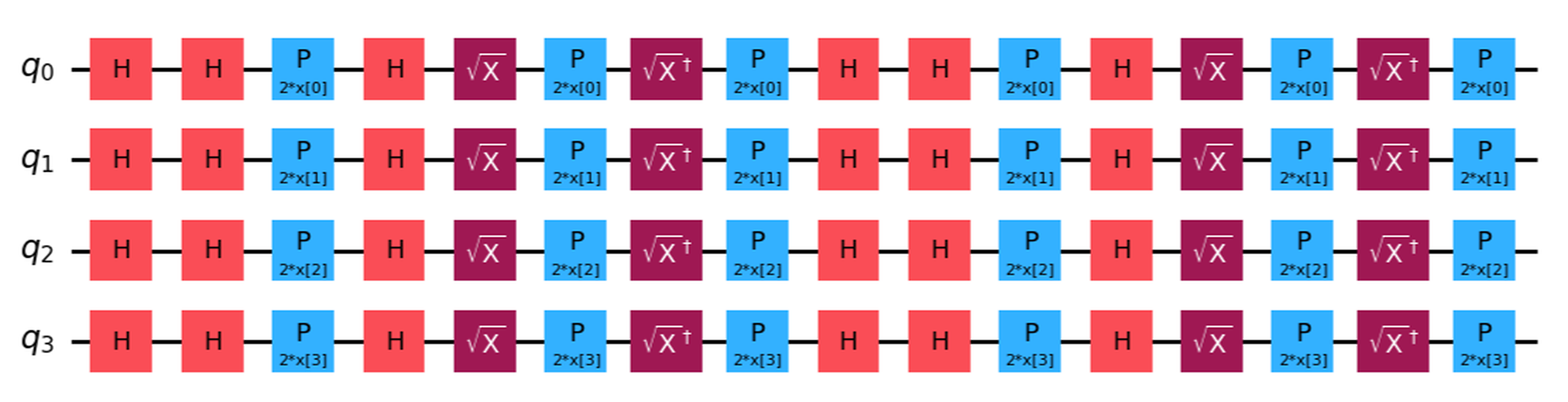}
    }
    \caption{The PauliFeatureMap in VQC-2 uses exponentials of Pauli operators to encode classical data, which facilitates transformations across multiple axes.}
    \label{fig:vqc2-feature-map}

\end{figure}

PauliFeatureMap encoding (Figure~\ref{fig:vqc2-feature-map}) can be expressed by the following equation, where $d$ is the number of repetitions, and $n$ is the number of qubits:

\[
U_{Pauli}(x) = \left( \exp\left(i \sum_{k} \phi_k(x) P_k\right) H^{\otimes n} \right)^d
\]

Here, $P_k$ represents selected Pauli operators or tensor products of Pauli operators. Due to incorporating multiple Pauli interactions the diversity of quantum states generated increases. Compared to the ZZFeatureMap, this encoding method produces richer representations.

\begin{figure}[H]
    \centering
    \includegraphics[width=\figwidth{65}]{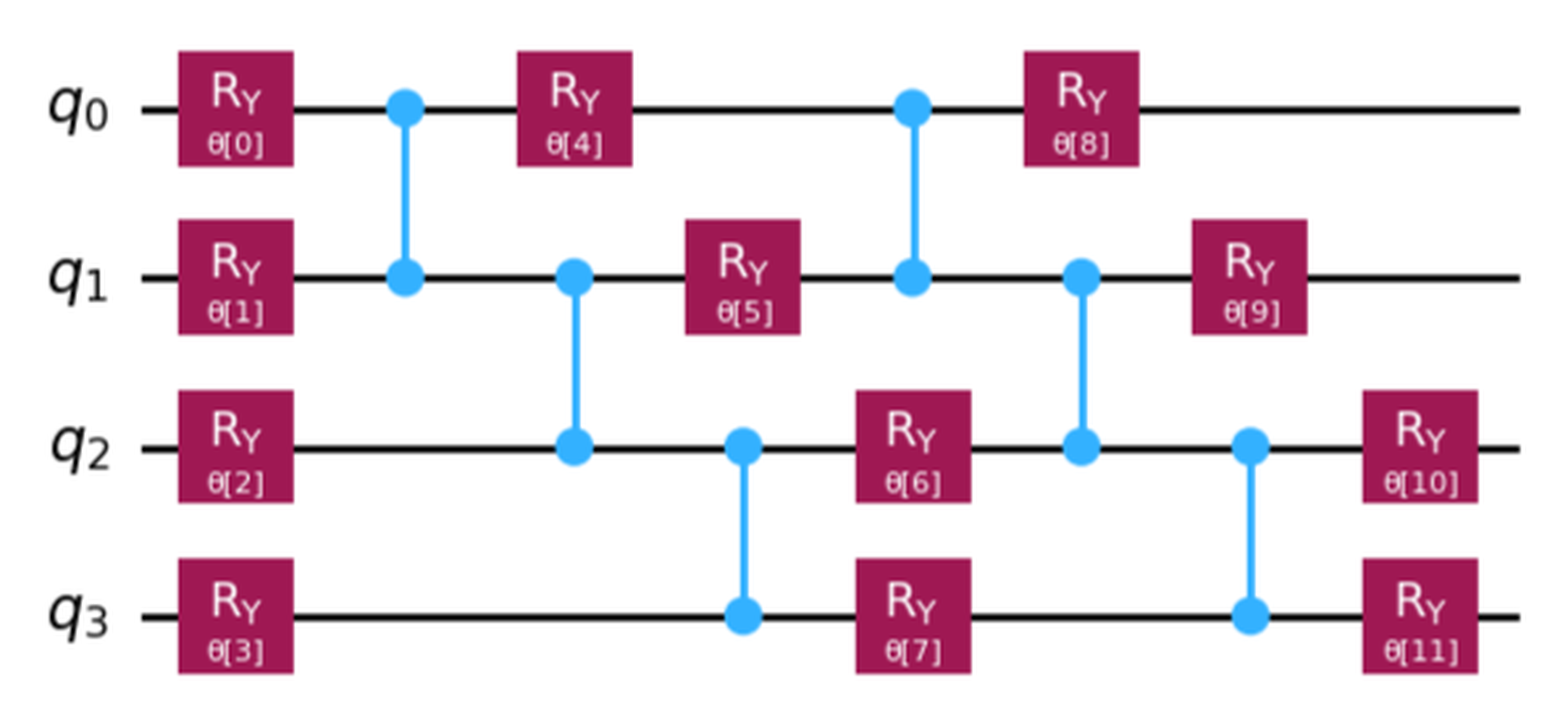}
    \caption{VQC-2 uses the TwoLocal ansatz circuit. It generalizes the structure of variational circuits by allowing configurable rotation gates and entanglement patterns.}
    \label{fig:vqc2-ansatz}

\end{figure}

In the TwoLocal circuit shown in Figure~\ref{fig:vqc2-ansatz}, $R_y$ gates are used for rotation, and Controlled-Z ($CZ$) gates for entanglement. The unitary operation for it can be expressed as follows, where $L$ is the number of repetitions and $n$ is the number of qubits:

\[
U_{TL}(\theta)=\left(\prod_{l=1}^{L}\left(\bigotimes_{i=1}^{n}R_{y}(\theta_{i,l})U_{CZ}\right)\right)\bigotimes_{i=1}^{n}R_{y}(\theta_{i,0})
\]

The ansatz structure is flexible and expressive, which enables modeling of more complex decision boundaries. However, the optimization landscape becomes more challenging.

VQC-2 utilizes a SamplerQNN, where outputs are derived from measurement probability distributions:

\[
p(y|x, \theta) = |\langle y | U(\theta) |\psi(x)\rangle|^2
\]

The use of SamplerQNN is motivated by the increased expressivity of the PauliFeatureMap and TwoLocal ansatz, which produce richer quantum states. Sampling-based outputs allow direct modeling of class probabilities.

The optimizer used is SPSA, chosen for its ability to efficiently handle the increased number of parameters without requiring explicit gradient computation.

\subsubsection{VQC-3: ZFeatureMap + RealAmplitudes}

VQC-3 investigates the effect of simplified feature encoding combined with a deeper variational circuit.

\begin{figure}[H]
\centering
\includegraphics[width=\figwidth{35}]{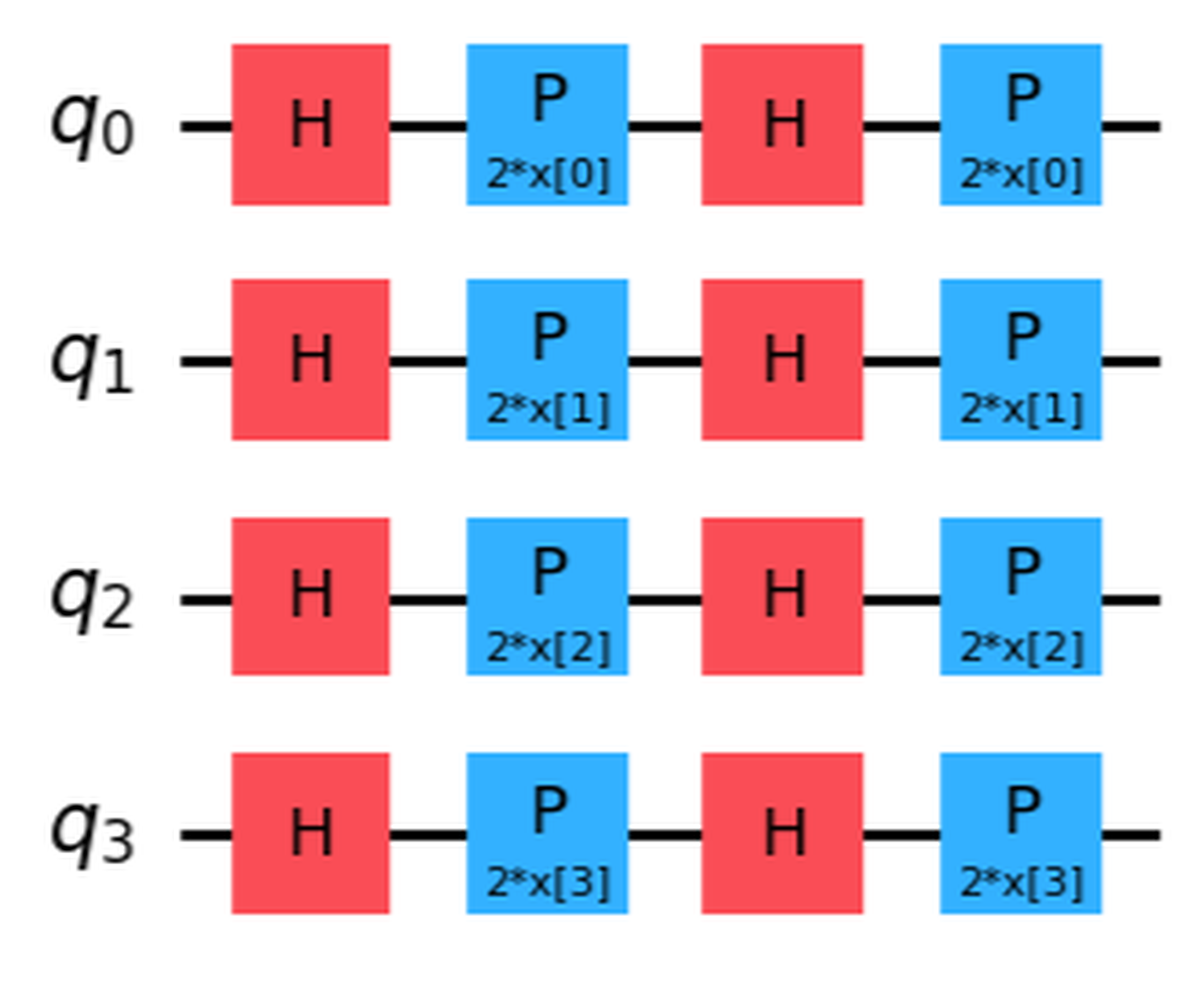}
\caption{VQC-3 uses the ZFeatureMap which performs encoding using only single-qubit Z-rotations.}
\label{fig:vqc3-feature-map}
\end{figure}

ZFeatureMap encoding (Figure~\ref{fig:vqc3-feature-map}) can be expressed as follows, where $d$ is the number of repetitions and $n$ is the number of qubits:

\[
U_{Z}(x) = \left( \exp\left(i \sum_{i=1}^n \phi_i(x) Z_i\right) H^{\otimes n} \right)^d
\]

This approach does not introduce entanglement during the encoding phase, resulting in a separable and low-complexity feature encoding without explicit inter-qubit correlations. Consequently, the burden of learning complex relationships shifts to the variational circuit.

\begin{figure}[H]
\centering
\includegraphics[width=\figwidth{75}]{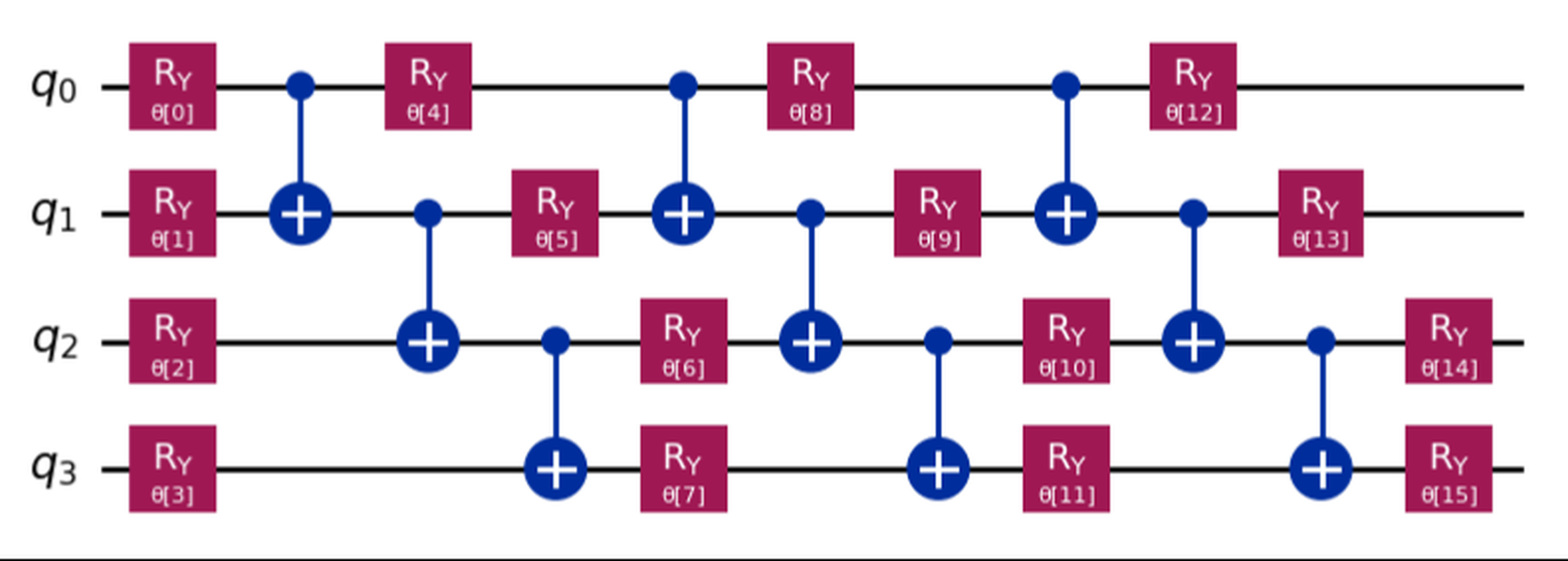}
\caption{VQC-3 uses a deeper RealAmplitudes ansatz with three repetitions.}
\label{fig:vqc3-ansatz}
\end{figure}

The deeper RealAmplitudes ansatz (reps = 3), shown in Figure~\ref{fig:vqc3-ansatz}, compensates for the lack of entanglement in the feature map. The increased depth enhances expressivity, allowing the circuit to model complex relationships despite the simpler encoding.

Due to the simple nature of the ZFeatureMap, EstimatorQNN is used for VQC-3. No entanglement is introduced in the feature encoding process, so the model depends on the ansatz for expressivity. The expectation values produced result in a more stable and continuous space.

VQC-3 employs the ADAM optimizer suitable for smoother optimization scenarios. It uses adaptive learning rates and momentum for better convergence.

\subsubsection{VQC-4: ZZFeatureMap + RealAmplitudes (Deep)}

VQC-4 is the most expressive model configuration in this study. It is a combination of deep feature encoding and a high capacity variational circuit.

\begin{figure}[H]
    \centering
    \makebox[\textwidth][c]{%
        \includegraphics[width=\figwidth{75}]{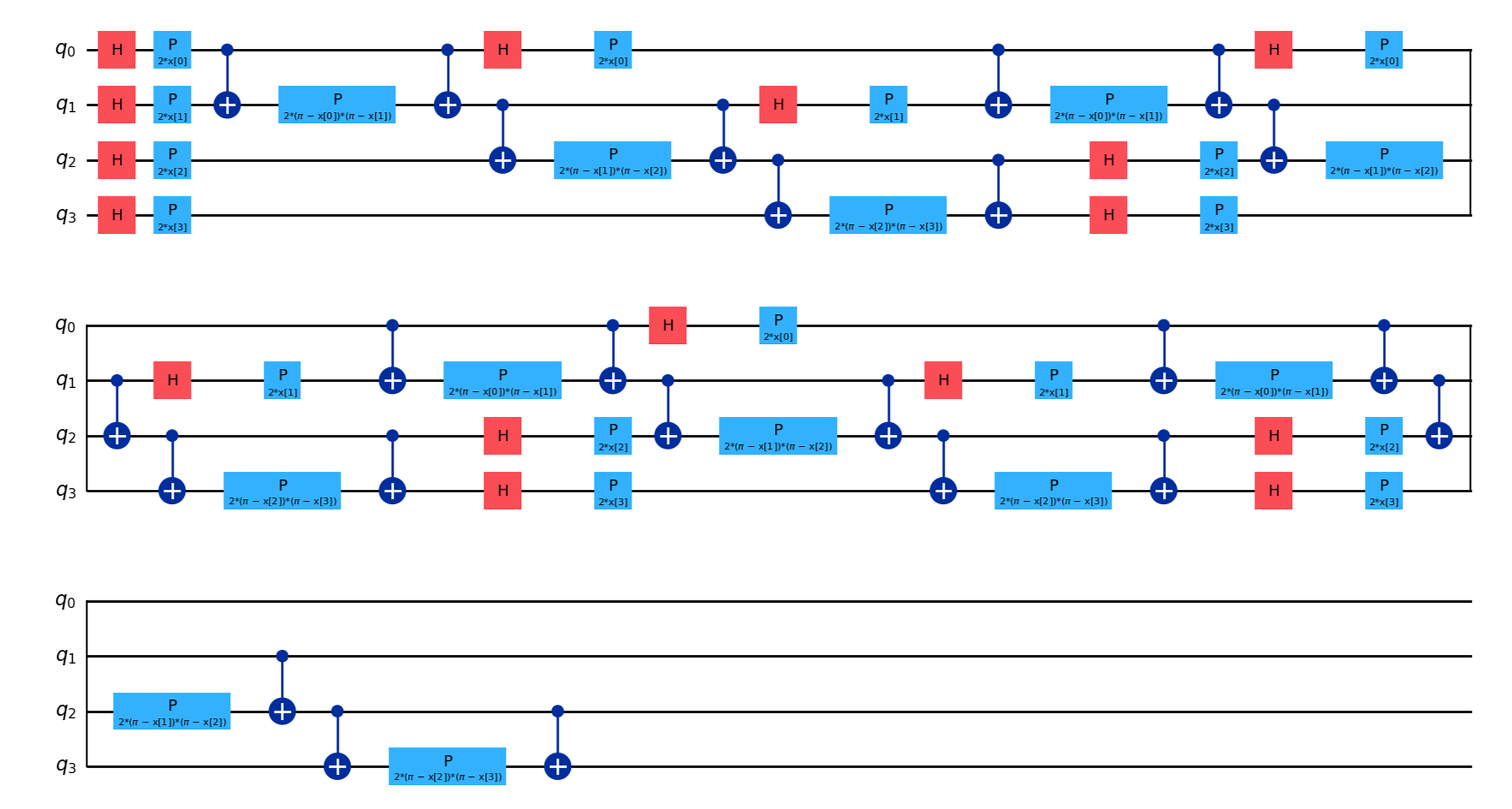}%
    }
    \caption{In VQC-4 the ZZFeatureMap is used with increased repetitions, which enhances the degree of entanglement and feature interaction.}
    \label{fig:vqc4-feature-map}

\end{figure}

In the ZZFeatureMap shown in Figure~\ref{fig:vqc4-feature-map}, the repeated application of the encoding layer (four reps) is intended to capture higher-order correlations in the data.

\begin{figure}[H]
    \centering
    \makebox[\textwidth][c]{
        \includegraphics[width=\figwidth{75}]{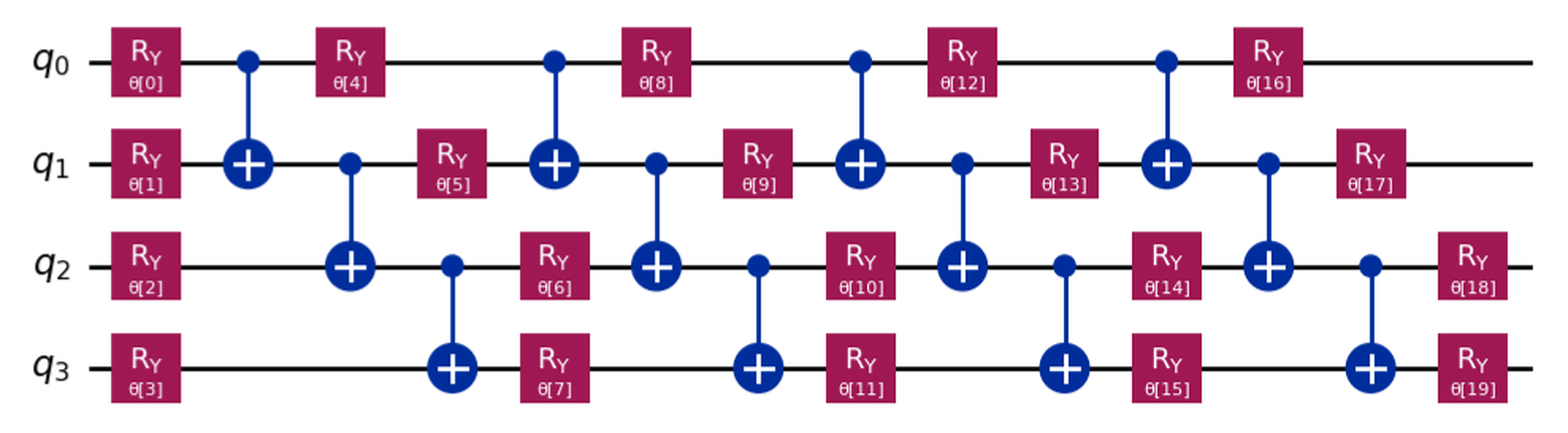}
    }
    \caption{VQC-4 is a deeper RealAmplitudes circuit with four repetitions.}
    \label{fig:vqc4-ansatz}

\end{figure}

The deeper ansatz shown in Figure~\ref{fig:vqc4-ansatz}, with four reps, is aimed at enabling the model to approximate highly complex functions. As the number of trainable parameters increases, the circuit expressivity increases.

EstimatorQNN is used for VQC-4. Although it has the most expressive circuit, deeper ZZFeatureMap and ansatz introduce significant complexity and potential noise. Having expectation values helps mitigate this additional stochasticity during the optimization process.

The COBYLA optimizer is used due to the increased complexity of the optimization process. It is a gradient-free method suitable for problems where gradient estimation becomes unreliable due to high-dimensions and non-convexity.

\subsection{Classical Benchmark Models}

Support Vector Machine (SVM) and Random Forest (RF) are two classical machine learning models trained alongside the proposed VQC architectures on the same synthetic data. These are benchmark classifiers for comparative evaluation. They were chosen in particular due to their known effectiveness in supervised anomaly detection and their ability to capture non-linear decision boundaries.

The SVM classifier was trained with a radial basis function (RBF) kernel. For high dimensional classification problems, SVM models have established effective results and strong generalization capability. Therefore, it serves as an appropriate baseline for evaluating the representational effectiveness of the VQC architectures in this study.

The Random Forest classifier uses ensemble learning techniques and is robust to overfitting. Output from multiple decision trees are aggregated in an ensemble of sub-models to generate the final answer. The model captures complex feature interactions and non-linear relationships in the data. It is chosen as a benchmark due to its stability and high classification performance in tabular data for cybersecurity and anomaly detection tasks.

A fair comparative analysis was ensured by using the same dataset, preproccessing pipeline, and train-test split ratio for both the classical benchmarks and VQC models.

\subsection{Evaluation Metrics}

This study concerns binary classification between normal and anomalous network traffic data. The evaluation metrics used include accuracy, precision, recall, and F1-score. Alongside overall predictive accuracy, the metrics also inform the quality of anomaly identification.

Accuracy is a measure of correct classifications among all predictions, and provides an overall performance indication. It is defined as:

\[
\text{Accuracy} = \frac{TP + TN}{TP + TN + FP + FN}
\]

where $TP$ represents true positives, $TN$ denotes true negatives, $FP$ represents false positives, and $FN$ denotes false negatives.

Precision measures detection performance for a particular class. It is calculated by taking the proportion of true positives to the total number of predictions in that specific class. High precision indicates a low false positive rate and reliable class type detection. Precision for each of the classes (normal and anomalous) is calculated according to the following formula and then macro-averaged.

\[
\text{Precision} = \frac{TP}{TP + FP}
\]

Recall measures the proportion of actual samples of a particular class that are correctly identified by a model. It is calculated by taking the proportion of true positives to the total number of data points in that class. Recall for each of the classes (normal and anomalous) is calculated according to the following formula and then macro-averaged.

\[
\text{Recall} = \frac{TP}{TP + FN}
\]

The F1-score is a harmonic mean of precision and recall. As it provides a balanced measure for both precision and recall, it is useful in scenarios where both false positives and false negatives are important considerations. The F1-score is defined as follows:

\[
F1\text{-Score} = 2 \times \frac{\text{Precision} \times \text{Recall}}{\text{Precision} + \text{Recall}}
\]

In addition to the qualitative metrics described above, confusion matrices and training loss curves were also visualized for evaluating model characteristics such as, classification behavior, convergence, and robustness.

\subsection{Experimental Setup}

All experiments were conducted in a python scientific computing setup utilizing Qiskit and additional machine learning libraries. Quantum circuits were executed using Qiskit Aer statevector simulation backends. In these simulations exact quantum state evolution and expectation values are computed, therefore, random perturbations due to noise are absent, unlike in real quantum hardware. Through this approach the experimental analysis isolates the study to understanding representational capabilities of the different VQC architectures and their optimization dynamics. The following table lists the primary software packages and dependencies used during experiment implementation.

\begin{table}[h]
\centering
\begin{tabular}{lc}
\hline
Software Package & Version \\
\hline
Python & 3.11.9 \\
Qiskit & 2.3.0 \\
Qiskit Aer & 0.17.2 \\
Qiskit Machine Learning & 0.9.0 \\
\hline 
\end{tabular} 
\caption{Software Environment Configuration}
\label{tab:software-environment-configuration}
\end{table}

\section{Experimental Results Analysis}

\subsection{Baseline VQC Performance}

The baseline experiment was conducted using preset variable parameters for the four Variational Quantum Classifier (VQC) models inorder to establish a reference point for sensitivity analysis. The results show that there are clear performance differences among the models, which is expected due to different combinations of feature encoding, ansatz, circuit depth, QNN, and optimizer types.

\begin{table}[h]
\centering
\begin{tabular}{lcccc}
\hline
Parameter Name & Value \\
\hline
Random Seed & 42 \\
Number of Samples & 1000 \\
Anomaly Ratio & 0.5 \\
Test Set Size & 0.3 \\
Max Optimizer Iterations & 100 \\
\hline 
\end{tabular} 
\caption{Baseline Parameters}
\label{tab:baseline-parameters}
\end{table}

\begin{table}[h]
\centering
% \resizebox{\linewidth}{!}{
\makebox[\textwidth][c]{
\begin{tabular}{lcccc}
\hline
Model & Accuracy & Precision & Recall & F1 Score \\
\hline
VQC-1 & 72.0 & 80.2 & 72.0 & 70.0 \\
VQC-2 & 57.3 & 77.0 & 57.3 & 47.8 \\
VQC-3 & 84.7 & 88.3 & 84.7 & 84.3 \\
VQC-4 & \textbf{89.0} & \textbf{89.8} & \textbf{89.0} & \textbf{88.9} \\
\hline 
\end{tabular}
}
\caption{Performance Summary of VQC Models}
\label{tab:performance}
\end{table}

\begin{figure}[H]
\centering
\includegraphics[width=\figwidth{65}]{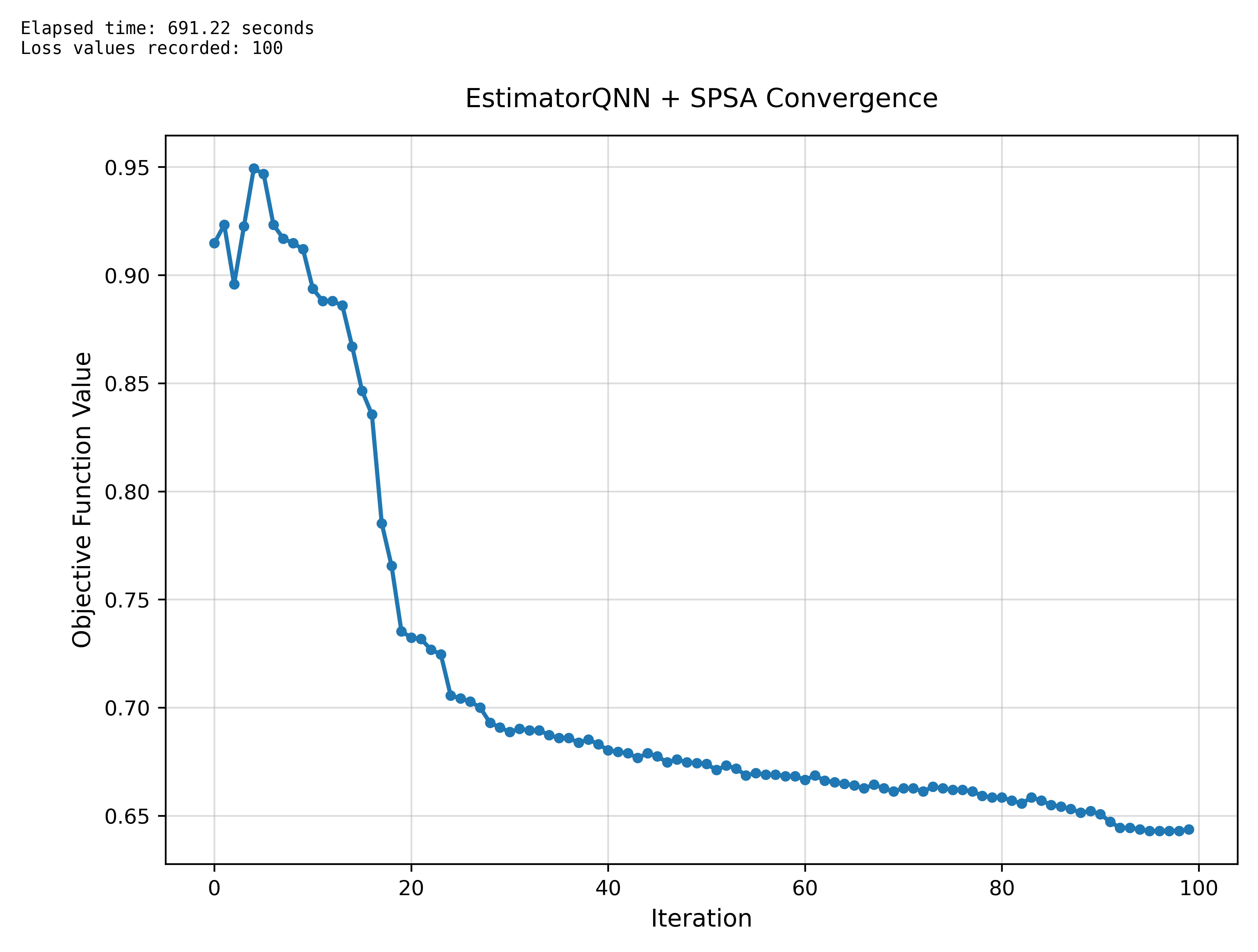}
\caption{Loss Curve of the first model optimized using SPSA, illustrates its optimization trajectory. }
\label{fig:vqc1-loss-curve}
\end{figure}

\begin{figure}[H]
\centering
\includegraphics[width=\figwidth{65}]{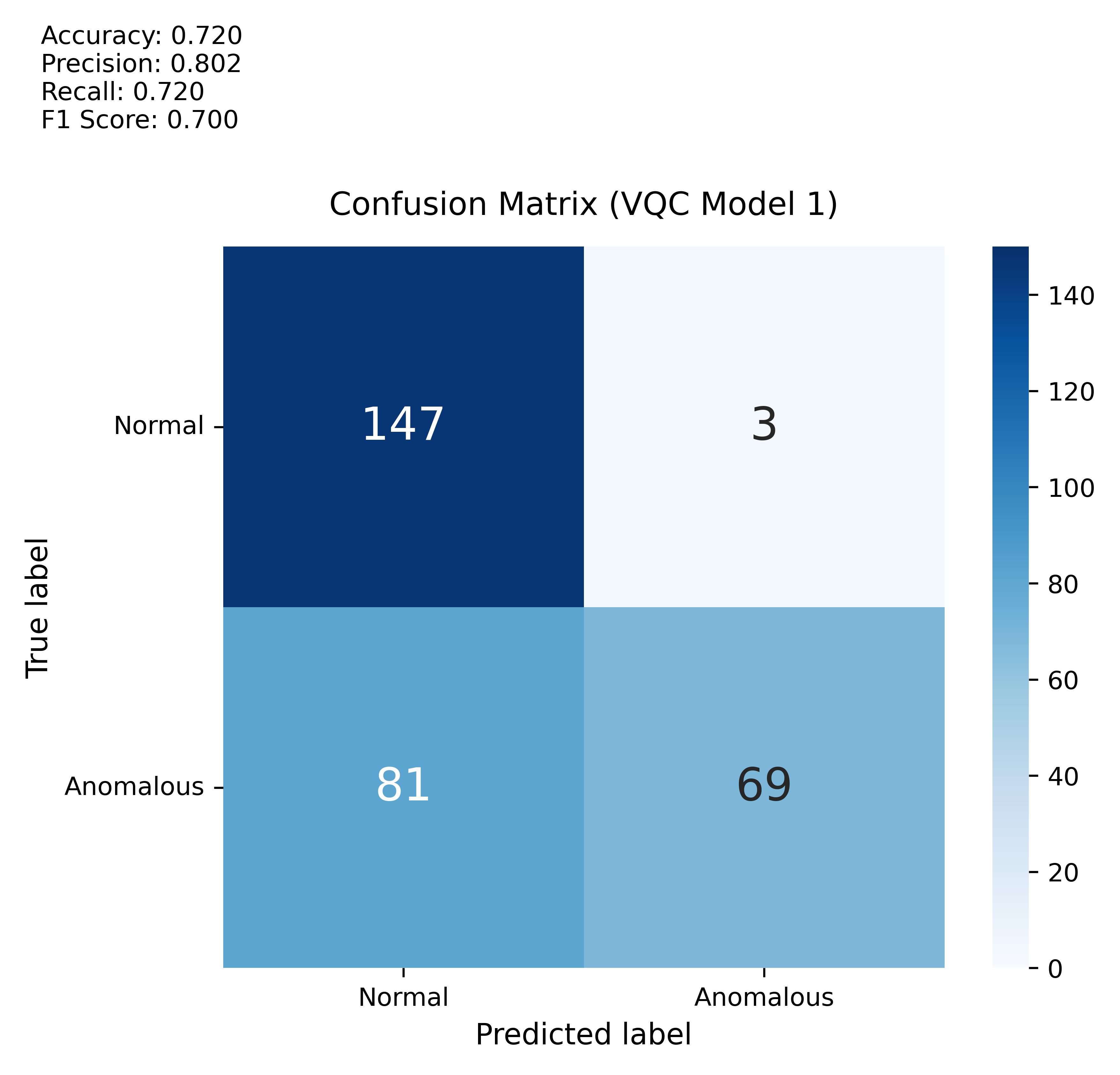}
\caption{Confusion Matrix of VQC-1. It achieved moderate performance with an accuracy of 72\% and a F1-Score of 70\%.}
\label{fig:vqc1-cm}
\end{figure}

The VQC-1 model demonstrated a high false negative rate (Figure~\ref{fig:vqc1-cm}), which indicates that anomaly detection capability of the model is weak. Although it had sufficient representational capability, the decision boundary for discriminating between the classes was not effectively learnt due to difficulty in optimization dynamics, with the corresponding loss curve shown in Figure~\ref{fig:vqc1-loss-curve}.

\begin{figure}[H]
\centering
\includegraphics[width=\figwidth{65}]{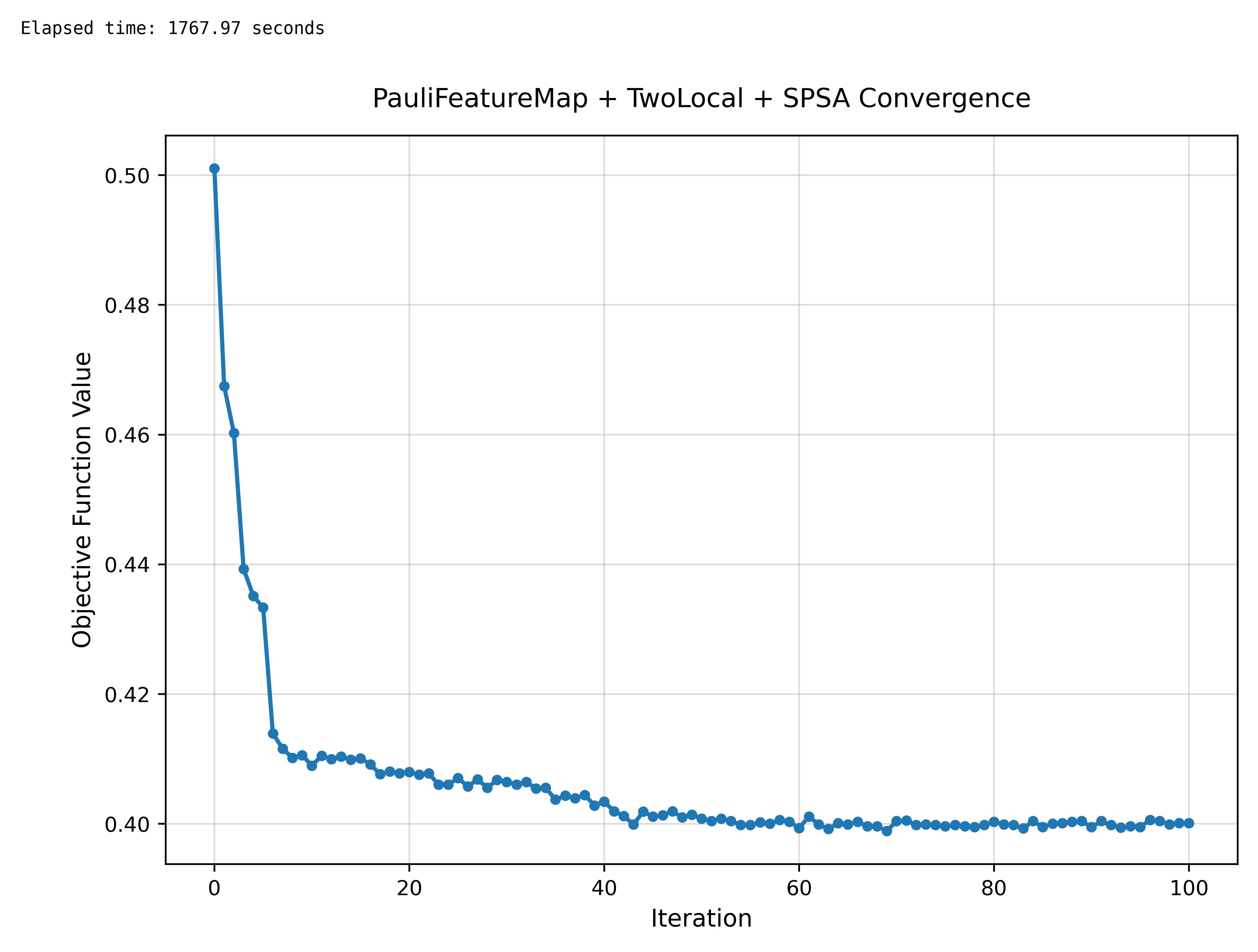}
\caption{The VQC-2 Loss Curve demonstrates its optimization trajectory.}
\label{fig:vqc2-loss-curve}
\end{figure}

\begin{figure}[H]
\centering
\includegraphics[width=\figwidth{60}]{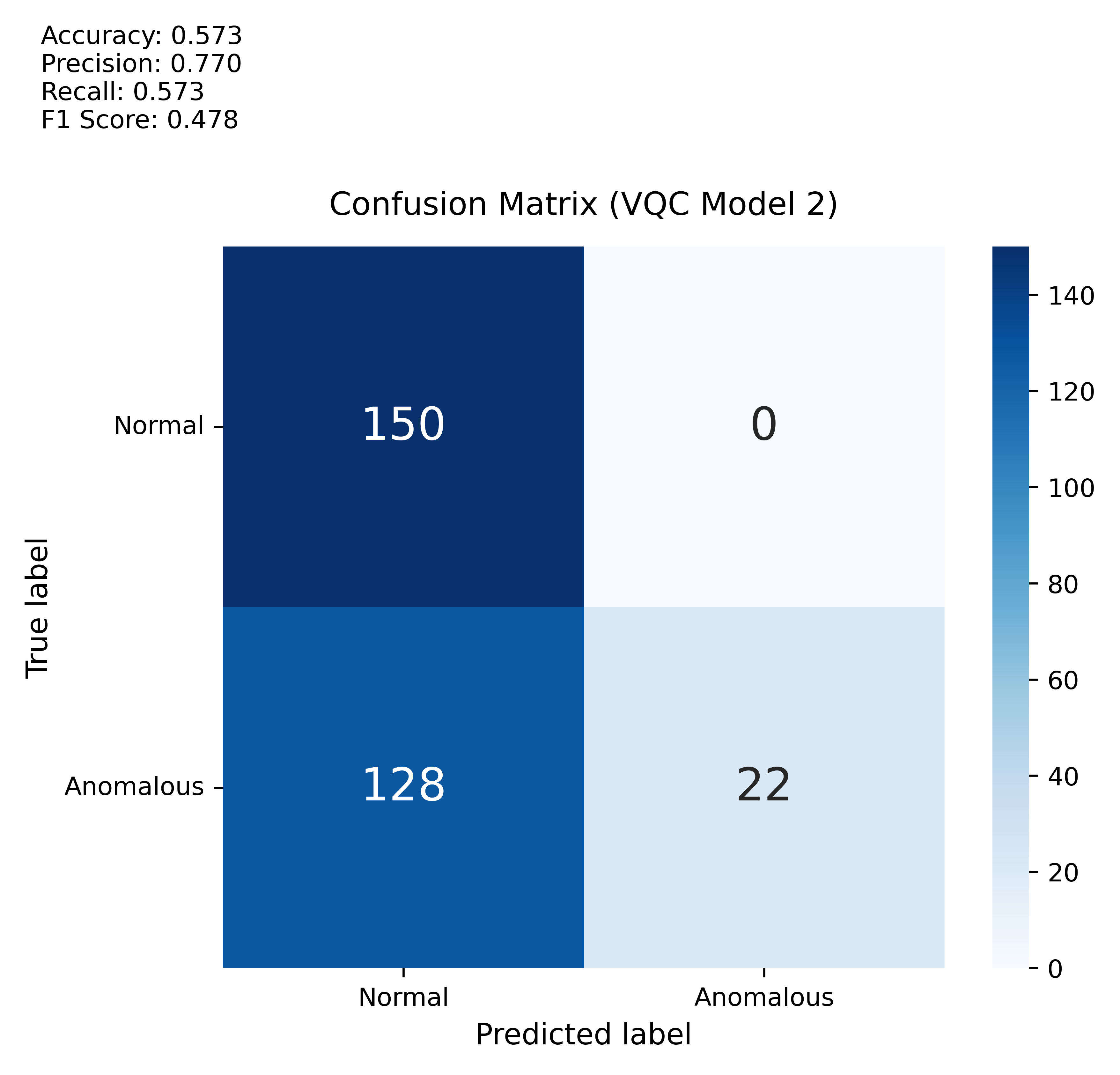}
\caption{VQC-2 evaluation metrics. It showed the weakest performance with an accuracy of approximately 57\% and a F1-Score of about 48\%.}
\label{fig:vqc2-cm}
\end{figure}

The VQC-2 loss curve in Figure~\ref{fig:vqc2-loss-curve} reaches a plateau after about 40 optimizer iterations, where the training loss converged to a value of about 0.4. The model consistently produced false negatives, as illustrated by the evaluation results in Figure~\ref{fig:vqc2-cm}.

\begin{figure}[H]
\centering
\includegraphics[width=\figwidth{60}]{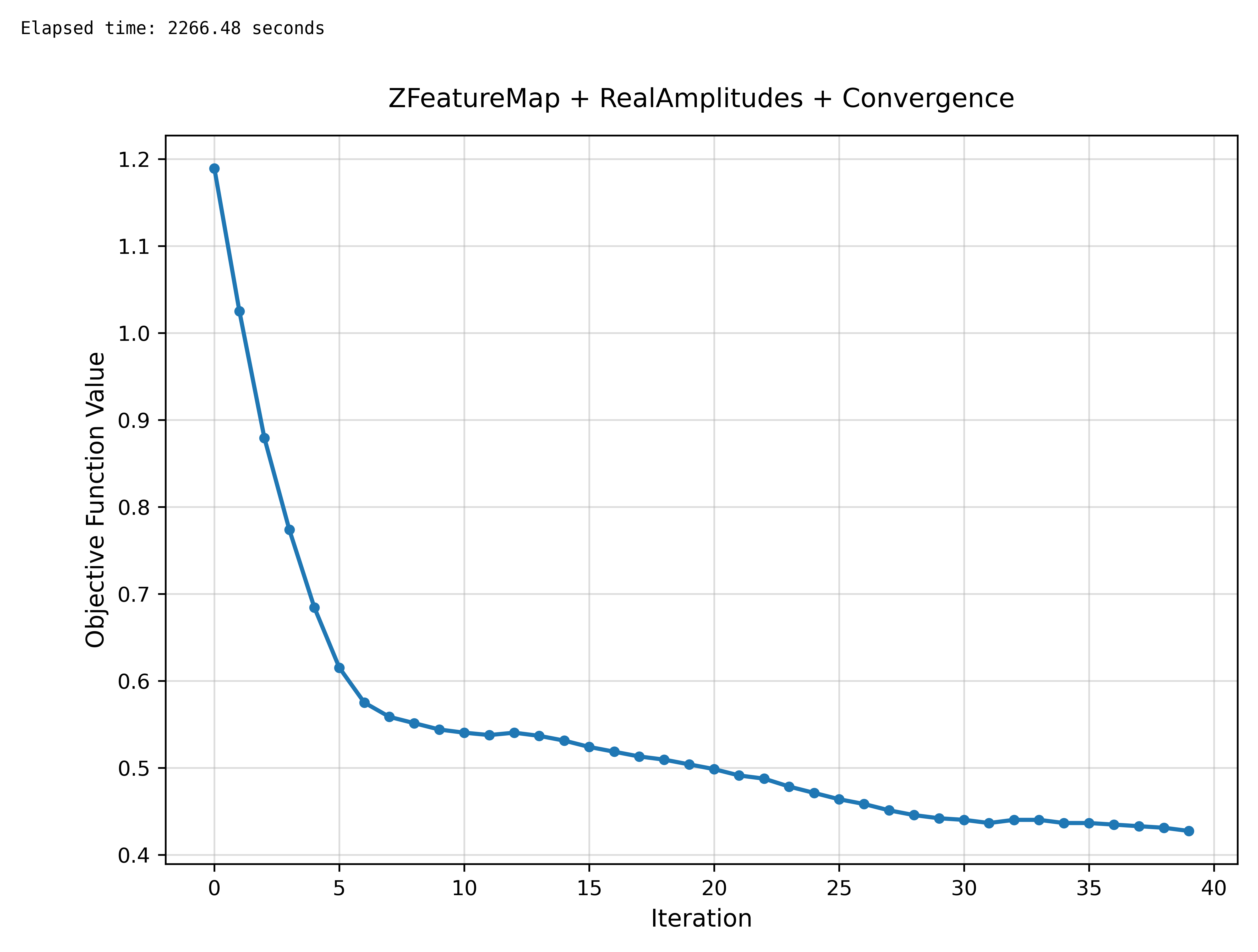}
\caption{The ADAM optimizer exhibited more stable convergence for VQC-3.}
\label{fig:vqc3-loss-curve}
\end{figure}

\begin{figure}[H]
\centering
\includegraphics[width=\figwidth{60}]{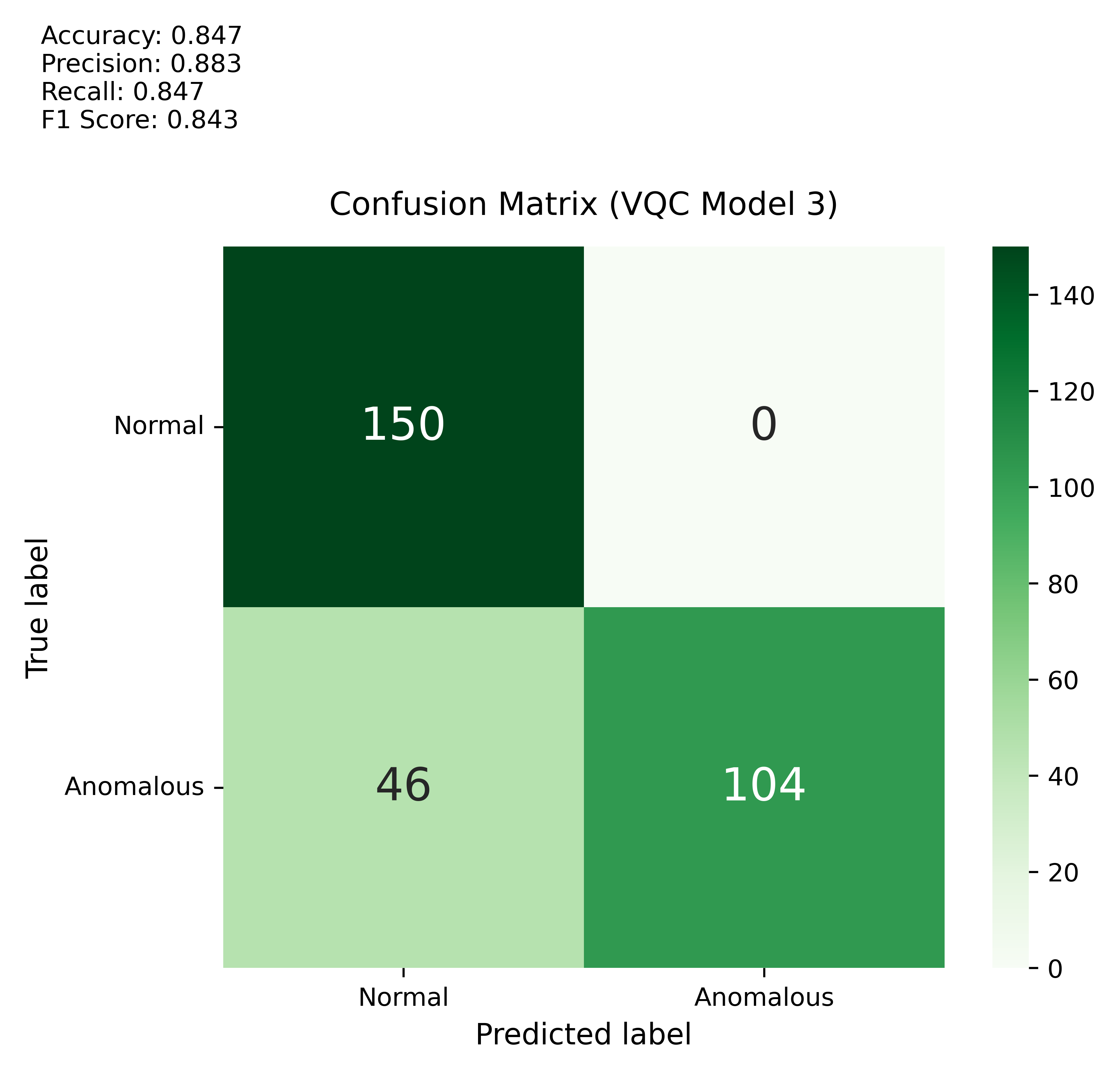}
\caption{VQC-3 evaluation metrics show substantial improvements over the previous models, achieving approximately 85\% accuracy and 84\% F1-Score.}
\label{fig:vqc3-cm}
\end{figure}

VQC-3 attained significantly lower false negatives and improved performance (Figure~\ref{fig:vqc3-cm}). The optimizer converges to a minima at about 40 iterations (Figure~\ref{fig:vqc3-loss-curve}). This indicates that the combination of simpler feature encoding and gradient-based optimization (ADAM) enabled the model to learn class boundaries more effectively.

\begin{figure}[H]
\centering
\includegraphics[width=\figwidth{60}]{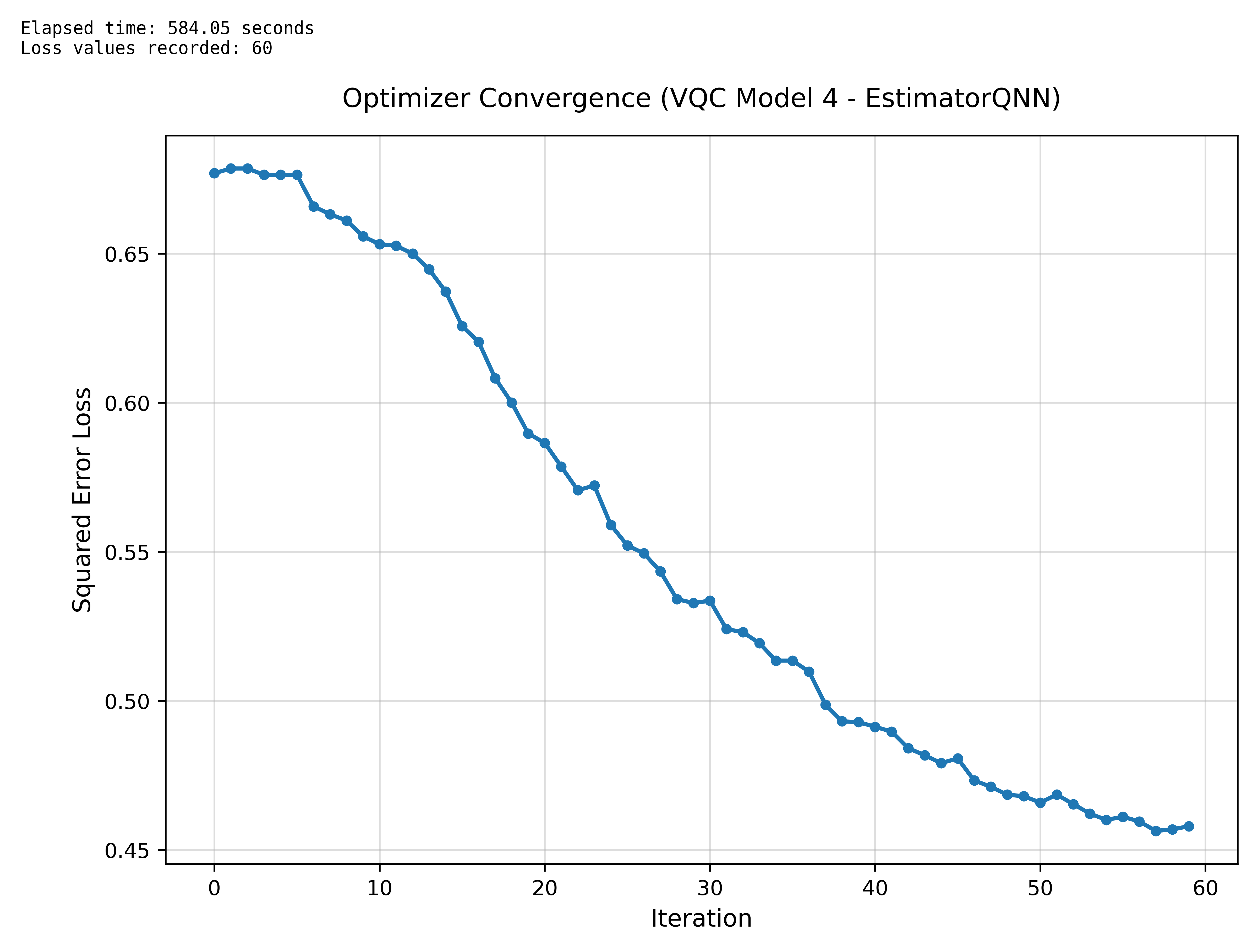}
\caption{Loss Curve of the deep VQC-4 model shows optimization using COBYLA had stable convergence.}
\label{fig:vqc4-loss-curve}
\end{figure}

\begin{figure}[H]
\centering
\includegraphics[width=\figwidth{60}]{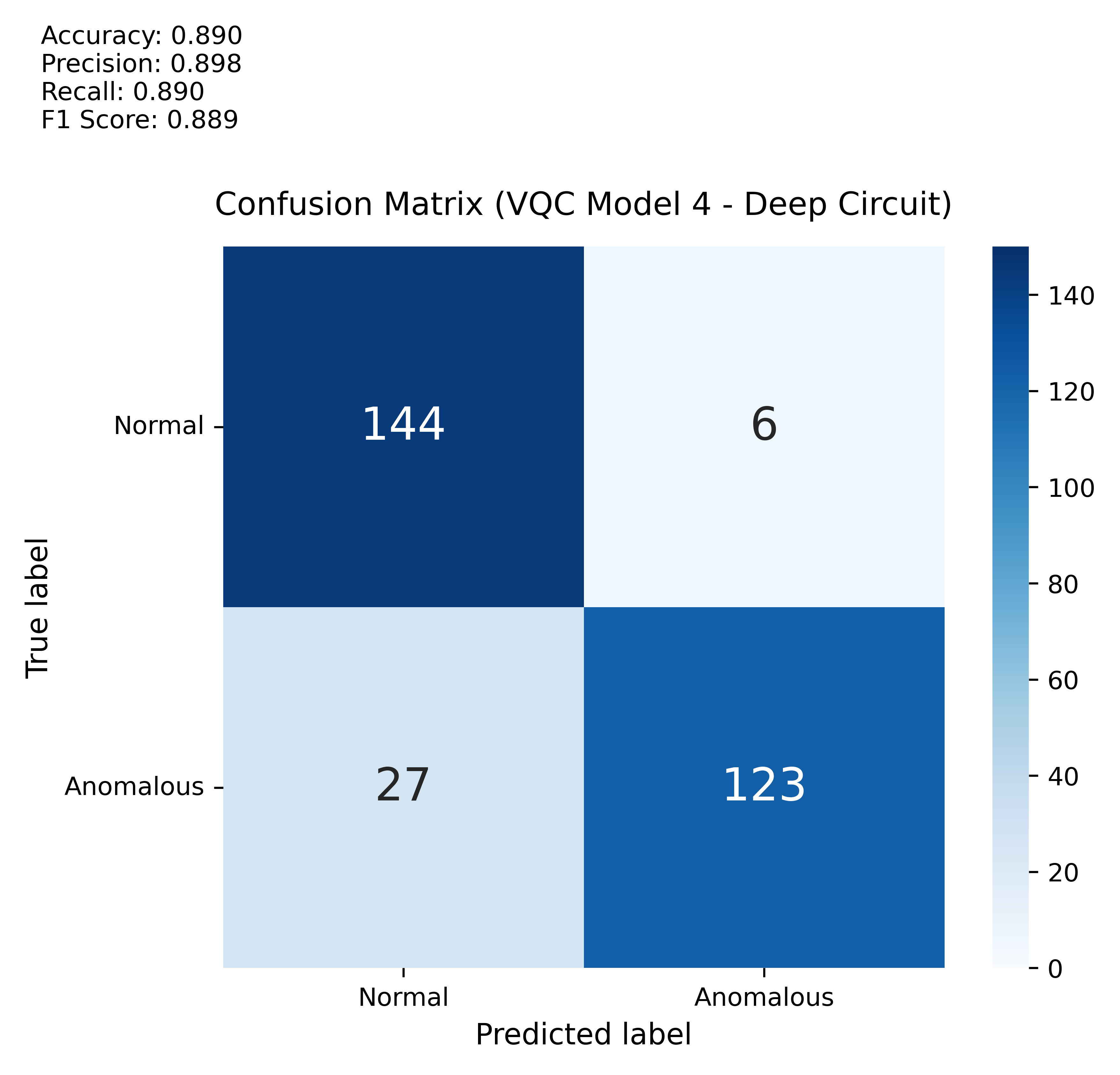}
\caption{VQC-4 evaluation metrics shows it achieved the strongest overall performance, reaching an accuracy of approximately 89\% and F1-Score of 89\%.}
\label{fig:vqc4-cm}
\end{figure}

VQC-4 has a deeper circuit architecture compared to VQC-1, and uses the COBYLA optimizer which converged at a minima at about 60 iterations (Figure~\ref{fig:vqc4-loss-curve}). The minima is slightly above 0.45 error value, lower in comparison to approximately 0.65 for VQC-1. Although there is a slight increase in false positives, a balanced classification is observed in Figure~\ref{fig:vqc4-cm}, implying a more effective decision boundary compared to the shallower models.

\subsection{Sensitivity Analysis}

We performed sensitivity analysis on VQC models to assess robustness, generalization, relative stability and adaptability to altered experimental conditions. The variables used in the analysis were the random seed used for initialization, size of dataset, anomaly ratio, train-test data partition ratio and the max number of iterations for the optimizer. Circuit configurations were kept the same.

\subsubsection{Sensitivity to Random Initialization}

The random seed was varied from the baseline value of 42 to 7 and 100 for investigating the effect of random initialization. For VQC-1, the accuracy remained close to the baseline value of approximately 72\% demonstrating stability across different initializations. VQC-2 has a similar trend with consistently weak anomaly detection performance, and high false negative rates reflected in the poor F1-scores.

In contrast, VQC-3 and VQC-4 showed noticeable sensitivity to initialization. Performance for VQC-3 improved under alternative random seeds in both metrics (accuracy and F1-Score), whereas performance of VQC-4 degraded in comparison to the baseline configuration. It can be inferred that although the deeper variational circuit has a higher representational capacity it is more susceptible to optimization instability as the non-convex loss landscapes may cause convergence into a higher local minima.

\begin{table}[htbp]
\centering
\begin{tabular}{lcccc}
\hline
\multicolumn{1}{c}{} & \multicolumn{2}{c}{Seed = 7} & \multicolumn{2}{c}{Seed = 100} \\ \hline
Model & Accuracy & F1-Score & Accuracy & F1-Score \\ \hline
VQC-1 & 72.7 & 70.7 & 72.7 & 70.6 \\
VQC-2 & 50.7 & 34.8 & 54.0 & 41.7 \\
VQC-3 & \textbf{87.7} & \textbf{87.5} & \textbf{91.0} & \textbf{90.9} \\
VQC-4 & 78.3 & 78.2 & 76.3 & 75.9 \\ \hline
\end{tabular}
\caption{Performance Under Different Random Seed Initializations}
\label{tab:random-seed-sensitivity}
\end{table}

\subsubsection{Impact of Dataset Size}

Model performance due to changes in the size of the dataset was also examined. Number of samples was taken once as 500 and again as 2000. For 500 samples, model performance increases compared to the 1000-sample baseline, due to a decrease in the size of the evaluation set. However, VQC-4 shows slightly reduced performance. For 2000 samples, due to an increase in the size of the evaluation set there is a decrease in classification performance for all models, although VQC-1 slightly improves. VQC-3 experiences a noticeable drop compared to the 500-sample run. VQC-4 performs the best in this increased data scenario (84.5\% accuracy, 84.4\% F1-Score).

\begin{table}[htbp]
\centering
\begin{tabular}{lcccc}
\hline
\multicolumn{1}{c}{} & \multicolumn{2}{c}{500 Samples} & \multicolumn{2}{c}{2000 Samples} \\ \hline 
Model & Accuracy & F1-Score & Accuracy & F1-Score \\ \hline
VQC-1 & 81.3 & 80.8 & 74.2 & 72.7 \\
VQC-2 & 59.3 & 51.3 & 53.7 & 41.0 \\
VQC-3 & \textbf{94.7} & \textbf{94.7} & 81.8 & 81.2 \\
VQC-4 & 86.0 & 86.0 & \textbf{84.5} & \textbf{84.4} \\ \hline
\end{tabular}
\caption{Performance Under Different Dataset Sizes}
\label{tab:dataset-size-sensitivity}
\end{table}

\subsubsection{Impact of Class Imbalance}

Anomaly ratio in the synthetic dataset was varied to 0.3 and 0.7 to observe the effect of class imbalance. In the case of lower anomaly ratio (0.3), VQC-1 obtained higher overall accuracy but lower F1-score compared to the baseline. VQC-2 biased heavily towards the majority-class and misclassified almost all anomalous samples. VQC-4 decreased accuracy and F1-score as well. In contrast, VQC-3 maintained robust classification performance with quite low false negatives.

In the case of a higher anomaly ratio (0.7), VQC-1 demonstrated improved anomaly detection performance. The imbalance caused VQC-2 to exhibit reversed classification bias, where now anomalous samples were correctly identified along with a large number of false positives. VQC-4 degraded substantially across all metrics. Similar to the low anomaly ratio case, VQC-3 again maintained near perfect classification performance.

\begin{table}[htbp]
\centering
\begin{tabular}{lcccc}
\hline
\multicolumn{1}{c}{} & \multicolumn{2}{c}{Ratio = 0.3} & \multicolumn{2}{c}{Ratio = 0.7} \\ \hline
Model & Accuracy & F1-Score & Accuracy & F1-Score \\ \hline
VQC-1 & 77.0 & 66.7 & 77.3 & 75.6 \\
VQC-2 & 70.0 & 41.2 & 74.7 & 55.8 \\
VQC-3 & \textbf{98.0} & \textbf{97.6} & \textbf{98.7} & \textbf{98.4} \\
VQC-4 & 81.3 & 74.2 & 71.7 & 69.7 \\ \hline
\end{tabular}
\caption{Performance Under Different Anomaly Ratios}
\label{tab:anomaly-ratio-sensitivity}
\end{table}

\subsubsection{Impact of Train-Test Split}

When the test size is reduced to 0.2, VQC-1 and VQC-2 experience decreases in both accuracy and F1-Score. VQC-4 exhibits a noticeable drop in performance compared to the baseline. VQC-3 improves above baseline, and obtains the highest accuracy (88.0\%) and F1-score (87.8\%).

When the test size is increased to 0.4, performance generally declines for all models due to a larger evaluation split. VQC-1 and VQC-2 experience decreases in both accuracy and F1-score. VQC-4 has a considerable drop in performance, indicating sensitivity to training-testing balance. VQC-3 remained the most robust model, and maintained high accuracy (86.8\%) and F1-score (86.5\%) near the baseline results.

\begin{table}[htbp]
\centering
\begin{tabular}{lcccc}
\hline
\multicolumn{1}{c}{} & \multicolumn{2}{c}{Test Size = 0.2} & \multicolumn{2}{c}{Test Size = 0.4} \\ \hline
Model & Accuracy & F1-Score & Accuracy & F1-Score \\ \hline
VQC-1 & 66.5 & 63.5 & 69.5 & 66.8 \\
VQC-2 & 55.5 & 44.5 & 51.7 & 37.1 \\
VQC-3 & \textbf{88.0} & \textbf{87.8} & \textbf{86.8} & \textbf{86.5} \\
VQC-4 & 83.5 & 83.3 & 77.2 & 76.6 \\ \hline
\end{tabular}
\caption{Performance Under Different Train-Test Split Ratios}
\label{tab:test-ratio-sensitivity}
\end{table}

\subsubsection{Impact of Optimization Iterations}

When the maximum optimizer iterations are reduced to 50, most models show a decline in performance. An obvious explanation being insufficient training. VQC-1 and VQC-4 have significant drops in the evaluation metrics. VQC-2 remains the weakest performing model. VQC-3 maintained its performance (86.3\% accuracy, 86.1\% F1-score) near baseline results, which suggests that it converges faster.

When the maximum optimizer iterations are increased to 200, VQC-1 obtains performance similar to baseline. VQC-2 and VQC-4 degrades in performance. VQC-3 remains stable, robust, and obtains similar results to the 50 iteration test-run due to its early stopping mechanism.

\begin{table}[htbp]
\centering
\begin{tabular}{lcccc}
\hline
\multicolumn{1}{c}{} & \multicolumn{2}{c}{Iterations = 50} & \multicolumn{2}{c}{Iterations = 200} \\ \hline
Model & Accuracy & F1-Score & Accuracy & F1-Score \\ \hline
VQC-1 & 67.0 & 64.1 & 72.3 & 70.0 \\
VQC-2 & 57.7 & 48.4 & 56.0 & 45.4 \\
VQC-3 & \textbf{86.3} & \textbf{86.1} & \textbf{85.3} & \textbf{85.0} \\
VQC-4 & 74.3 & 74.2 & 79.3 & 78.7 \\ \hline
\end{tabular}
\caption{Performance Under Different Optimizer Iteration Limits}
\label{optimizer-iteration-sensitivity}
\end{table}

\subsection{Misclassification Analysis}

\begin{figure}[H]
\centering
\includegraphics[width=\figwidth{60}]{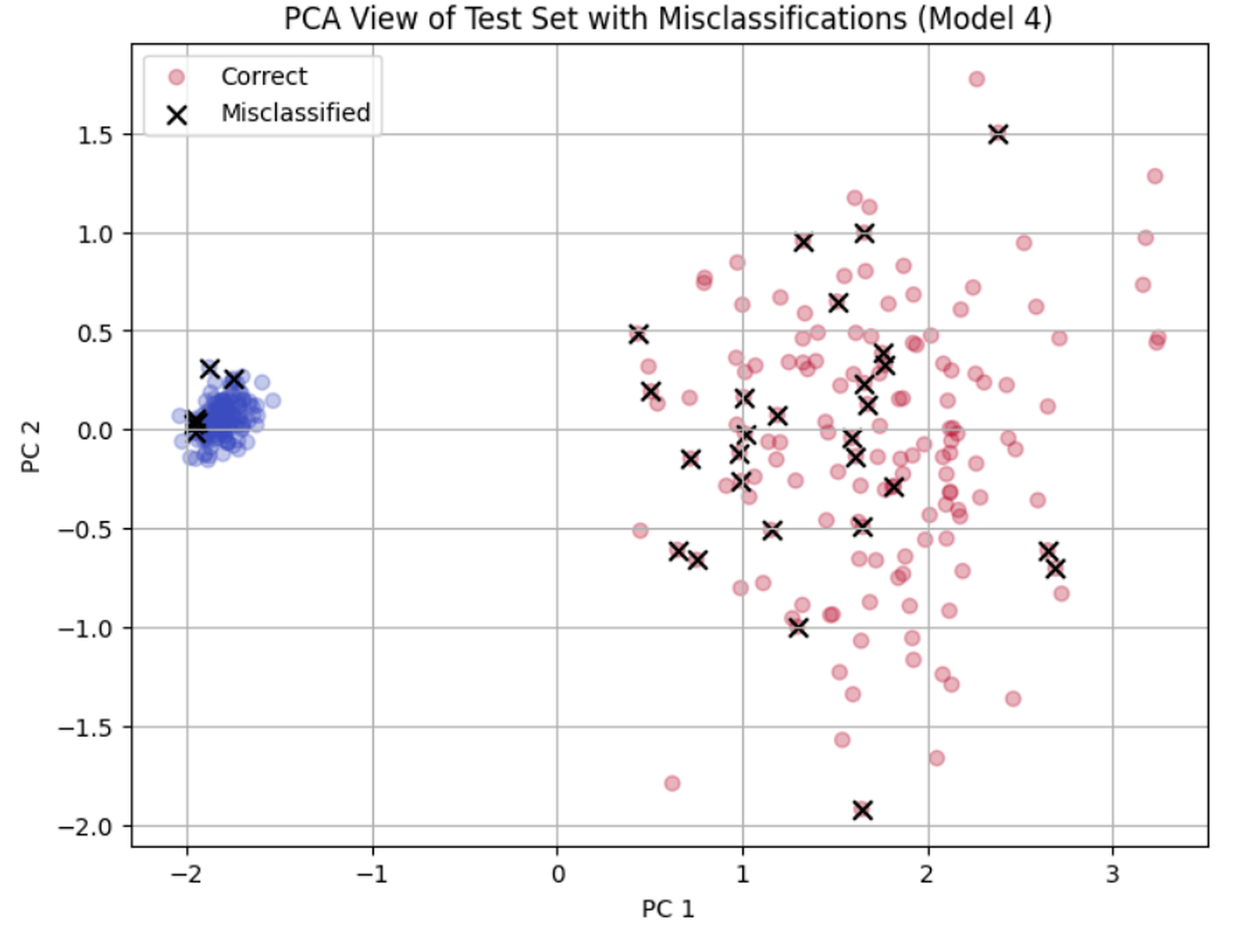}
\caption{Misclassification Analysis of VQC-4 by PCA visualization of data. Most misclassifications occur within the anomalous cluster.}
\label{fig:misclassification-pca}
\end{figure}

\begin{figure}[H]
\centering
\includegraphics[width=\figwidth{60}]{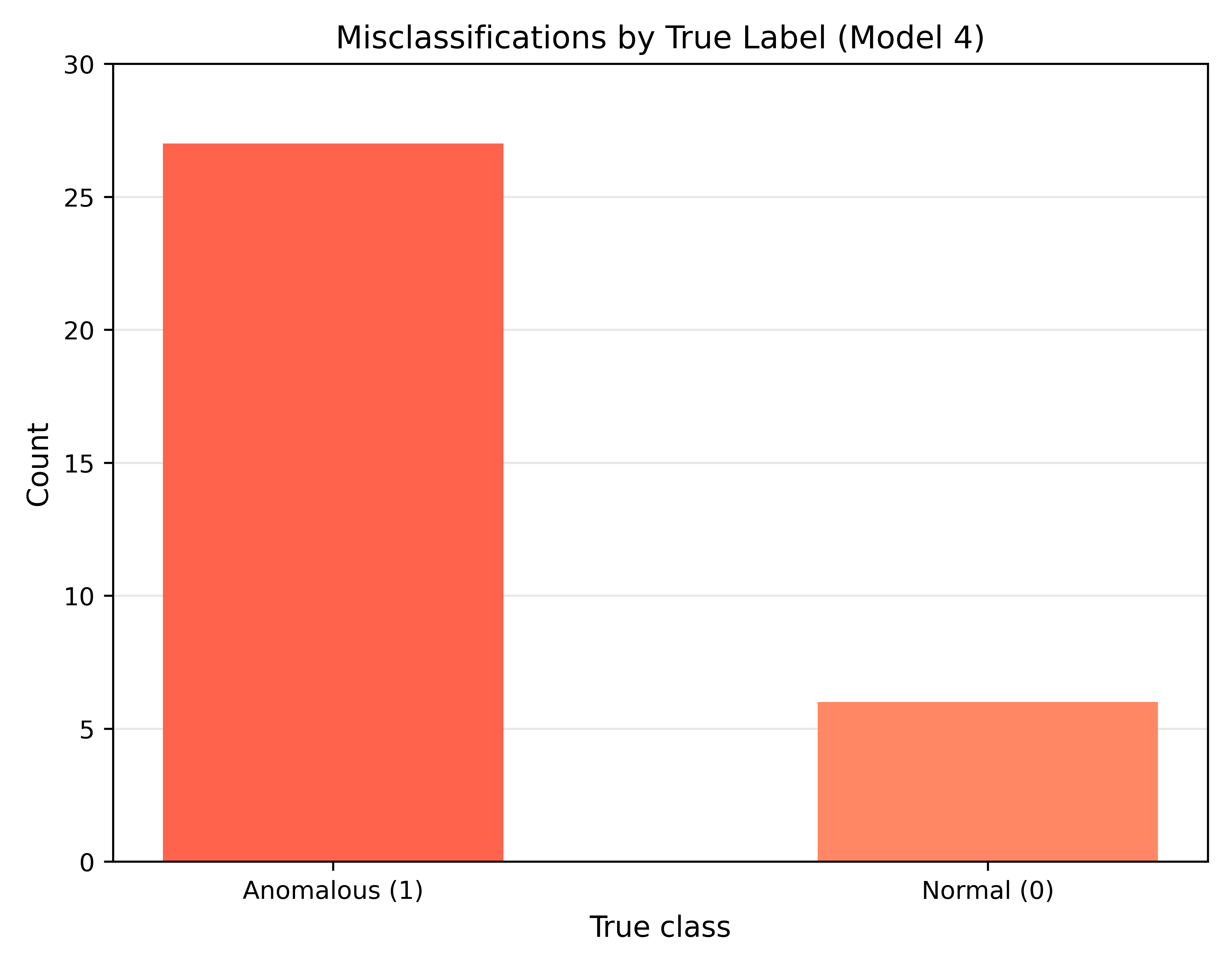}
\caption{VQC-4 misclassifications by True Label. The number of incorrectly classified anomalous samples is substantially higher than that of normal samples.}
\label{fig:misclassification-by-label}
\end{figure}

A detailed misclassification analysis was conducted for the VQC-4 model to further investigate the classification behavior of VQCs. Classified points were reduced to 2 dimensions using Principal Component Analysis (PCA). The scatter plot and class-wise error bar plot in Figures~\ref{fig:misclassification-pca} and~\ref{fig:misclassification-by-label}, respectively, provide insight about failure modes.

The VQC-4 model, for instance, captures the dominant structures of anomalous flows rather well; however, it is not as effective at learning the patterns of anomalous flows. As indicated in the evaluation results, class-wise normal samples remain tightly clustered and almost perfectly classified, hence, strong separability for that class. The class-wise misclassification counts support this observation, as only a few normal flows are misclassified. The rest of the classification challenges, however, are due to the variability and dispersion of anomalous network data.

\subsection{Comparison With Classical Models}

The VQC models studied in this work were compared to two of the most well-known and widely used classical machine learning models, namely Support Vector Machines (SVM) with an RBF kernel and Random Forest. The competitive evaluation results of the baseline experiments are summarized in the following table.

\begin{table}[h]
\centering
\begin{tabular}{lcccc}
\hline
Model & Accuracy & Precision & Recall & F1 Score \\
\hline
VQC-1 & 72.0 & 80.2 & 72.0 & 70.0 \\
VQC-2 & 57.3 & 77.0 & 57.3 & 47.8 \\
VQC-3 & 84.7 & 88.3 & 84.7 & 84.3 \\
VQC-4 (Deep) & 89.0 & 89.8 & 89.0 & 88.9 \\
SVM (RBF) & \textbf{100.0} & \textbf{100.0} & \textbf{100.0} & \textbf{100.0} \\
Random Forest & \textbf{100.0} & \textbf{100.0} & \textbf{100.0} & \textbf{100.0} \\
\hline 
\end{tabular} 
\caption{Performance Summary of Quantum and Classical Models}
\label{tab:performance-QML-CML}
\end{table}

\begin{figure}[H]
    \centering
    \makebox[\textwidth][c]{
        \includegraphics[width=\figwidth{75}]{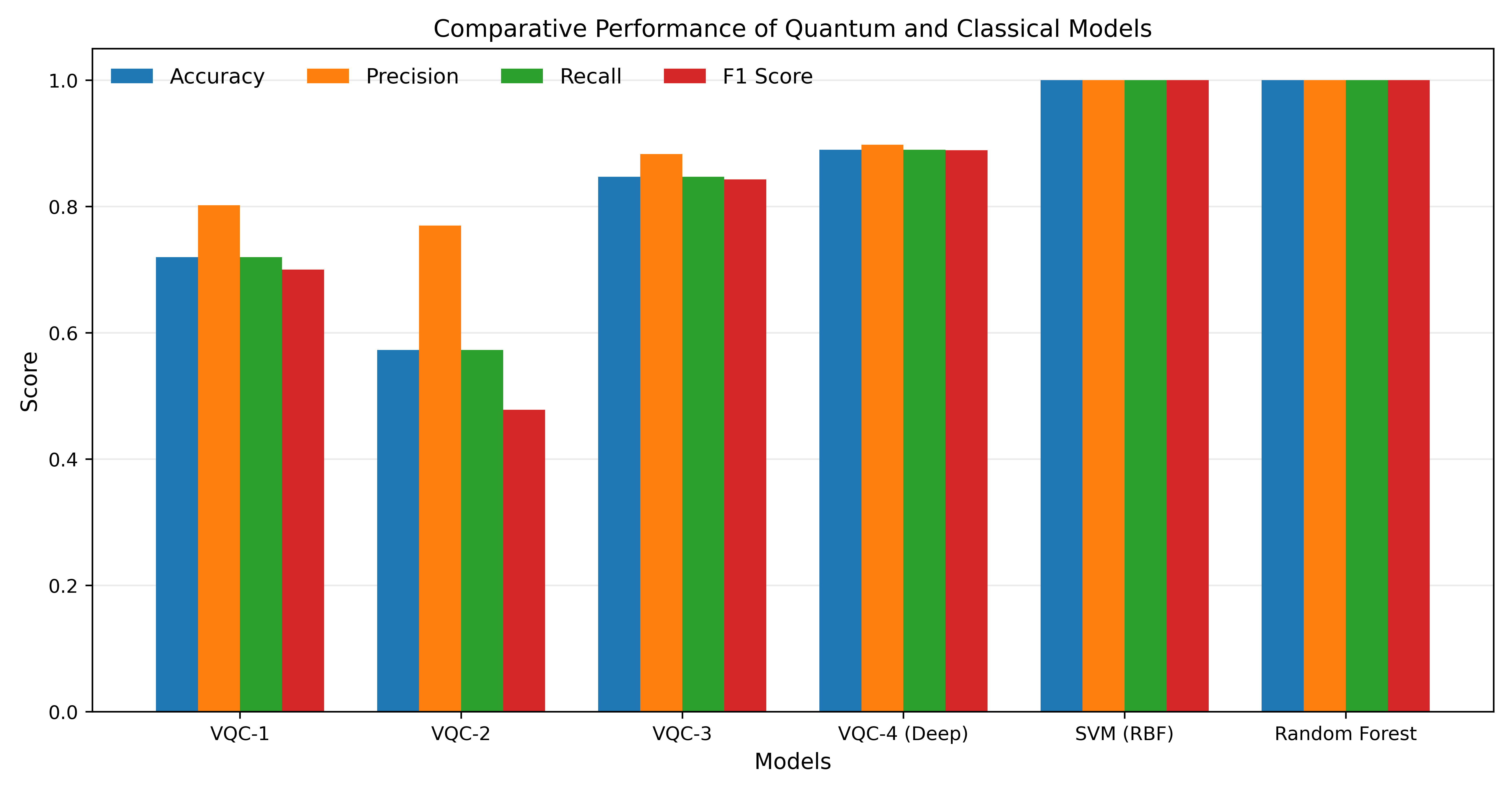}
    }
    \caption{Bar-chart of Comparative Performance of Quantum and Classical Models. The classical models substantially outperformed all evaluated VQC architectures, achieving perfect classification performance.}
    \label{fig:performance-comparison}

\end{figure}

As shown in Figure~\ref{fig:performance-comparison}, both Random Forest and SVM successfully separated normal and anomalous traffic instances. For classical machine learning algorithms the generated synthetic dataset is highly separable. The feature distribution and PCA plots presented in the dataset analysis section support this fact, because the data show clear clustering behavior and strong feature correlations.

\subsection{Discussion of Findings}

Consistent trends emerge from the experimental results with respect to the characteristics and behavior of variational quantum classifiers for the detection of anomalous network traffic. Among the four models evaluated, VQC-3 demonstrated the best stability, robustness, and convergence consistency. It is also observed that deeper variational circuits improve representational capability but introduce greater optimization instability. Although VQC-4 obtained the highest baseline performance, it was sensitive to variations in random initialization, dataset size, anomaly ratio, train-test ratio, and optimizer iterations. In contrast, the shallower VQC-1 maintained stable but comparatively weaker classification performance. VQC-2 obtained the weakest anomaly detection capability and frequently exhibited class-dependent bias resulting in a large number of misclassifications in the opposite class.

For multiple cases, VQC-1 and VQC-2 obtained moderate accuracy but at the same time produced a large number of false negatives. It can be inferred from this that accuracy alone is insufficient for evaluation of anomaly detection systems. The misclassification analysis of VQC-4 demonstrated that the majority of wrong predictions occurred within the anomalous cluster.

Benchmarking the quantum models against strong classical contenders like Random Forest and Support Vector Machine reveals a performance gap. The synthetic dataset in this study had low intrinsic complexity, thus allowing the classical ML models to create highly effective decision boundaries in order to obtain perfect classification. Thus for relatively structured datasets that are highly separable, classical machine learning remain substantially more efficient and reliable than current NISQ-era variational quantum classifiers.

\section{Conclusion \& Future Works}

\subsection{Conclusion}

This study explored multiple VQC architectures by varying feature encoding methods, ansatz structures, QNN type, and classical optimizers. The systematic evaluation was carried out in the context of network traffic anomaly detection using a synthetic dataset. Insights have been gained regarding the optimization dynamics and representational capacity of NISQ-era VQC models.

The experimental results showed that deeper circuits like VQC-4 achieved higher performance of upto 89\% accuracy. However, it had significant sensitivity to training conditions. In contrast, the shallower VQC-1 maintained relatively stable optimization but achieved a moderate accuracy of upto 72\% along with high false negative rates. VQC-2, composed of PauliFeatureMap and TwoLocal ansatz, exhibited the weakest overall performance. This model frequently demonstrated a severe class-dependent bias and inability to effectively model the data distribution. The best performing model, VQC-3, composed of simpler ZFeatureMap combined with a deeper RealAmplitudes ansatz (3 reps), and optimized using ADAM demonstrated strongest overall robustness and convergence behavior.

Support Vector Machine and Random Forest models substantially outperformed all evaluated VQC models by achieving perfect classification accuracy. However, the performance of VQC-3 and VQC-4 demonstrates that quantum models are capable of learning meaningful decision boundaries for classification, and can contend fairly despite having limited quantum resources and optimization constraints. The representational capabilities of these hybrid quantum-classical models are promising, and opportunities for quantum advantage should be explored by further investigations for increasingly higher-dimensional, noisier, or complex detection tasks where classical systems tend to under-perform.

\subsection{Future Work}

In order to enhance the efficacy of these quantum-based anomaly detectors, several future research directions may be pursued based on the empirical findings of this study. An important direction involves extending the current static features of the data to include time-series information. This would allow the quantum models to better capture the sequential and temporal dynamics found in real-world use cases. In addition, future work may explore hybrid quantum-classical ensemble models that combine predictive output from both detector types. This approach could leverage the expressive power of high-dimensional quantum feature spaces, along with the stability of classical optimization.

To address NISQ-era hardware constraints, subsequent research should apply circuit pruning techniques or adaptive ansatz construction methods to systematically minimize gate depth and mitigate hardware noise. Finally, implementing automatic hyperparameter optimization will be crucial to dynamically determine ideal circuit configurations.

\end{document}